\documentclass[journal]{IEEEtran}

\usepackage{acronym}
\usepackage{amsfonts}
\usepackage[dvips]{graphicx}
\usepackage{times}
\usepackage{cite}
\usepackage{amsmath}
\usepackage{array}
\usepackage{amssymb}
\usepackage{stfloats}
\usepackage{diagbox}
\usepackage{graphicx}
\usepackage{footnote}
\usepackage{amsthm}
\usepackage{booktabs}
\usepackage{array}
\usepackage[ruled,vlined]{algorithm2e}
\usepackage{subeqnarray}
\usepackage{cases}
\usepackage{threeparttable}
\usepackage{color}
\usepackage{epstopdf}
\usepackage{multirow}
\usepackage{tabularx}
\usepackage{enumerate}
\usepackage{multicol}
\usepackage{hyperref}
\usepackage{subfig}
\usepackage{caption}

\newtheorem{proposition}{Proposition}

\def\bb0{{\mathbb{0}}}

\def\ba{{\mathbf{a}}}
\def\bb{{\mathbf{b}}}

\def\bd{{\mathbf{d}}}

\def\bn{{\mathbf{n}}}

\def\br{{\mathbf{r}}}
\def\bs{{\mathbf{s}}}

\def\bu{{\mathbf{u}}}
\def\bv{{\mathbf{v}}}
\def\bw{{\mathbf{w}}}
\def\bx{{\mathbf{x}}}
\def\by{{\mathbf{y}}}
\def\bz{{\mathbf{z}}}
\def\b0{{\mathbf{0}}}

\def\bA{{\mathbf{A}}}

\def\bC{{\mathbf{C}}}
\def\bD{{\mathbf{D}}}
\def\bE{{\mathbf{E}}}
\def\bF{{\mathbf{F}}}
\def\bG{{\mathbf{G}}}
\def\bH{{\mathbf{H}}}
\def\bI{{\mathbf{I}}}

\def\bP{{\mathbf{P}}}
\def\bQ{{\mathbf{Q}}}

\def\bS{{\mathbf{S}}}

\def\bU{{\mathbf{U}}}
\def\bV{{\mathbf{V}}}
\def\bW{{\mathbf{W}}}
\def\bX{{\mathbf{X}}}
\def\bY{{\mathbf{Y}}}
\def\bZ{{\mathbf{Z}}}

\def\bbC{{\mathbb{C}}}

\def\cF{\mathcal{F}}
\def\cG{\mathcal{G}}

\def\cN{\mathcal{N}}
\def\cO{\mathcal{O}}
\def\cP{\mathcal{P}}
\def\cQ{\mathcal{Q}}
\def\cR{\mathcal{R}}

\def\sfE{\mathsf{E}}

\def\sfP{\mathsf{P}}

\def\sfX{\mathsf{X}}

\def\sfj{{\mathsf{j}}}

\def\sf0{{\mathsf{0}}}

\def\rmD{\mathrm{D}}

\def\sign{{\mathrm{sign}}}

\def\rmd{{\mathrm{d}}}

\def\rmx{{\mathrm{x}}}

\def\rm0{{\mathrm{0}}}

\def\Nt{{N_\mathrm{T}}}

\def\Nr{{N_\mathrm{R}}}

\def\Nc{{N_\mathrm{c}}}

\def\Nct{{N_{\mathrm{t}}}}
\def\Ncr{{N_{\mathrm{r}}}}
\def\Ns{{N_\mathrm{s}}}

\acrodef{CSI}[CSI]{channel state information}
\acrodef{CSIT}[CSIT]{channel state information at the transmitter}
\acrodef{CSIR}[CSIR]{channel state information at the receiver}
\acrodef{MIMO}[MIMO]{multiple-input multiple-output}
\acrodef{SISO}[SISO]{single-input single-output}
\acrodef{MISO}[MISO]{multiple-input single-output}
\acrodef{SIMO}[SIMO]{single-input multiple-output}
\acrodef{ADCs}[ADCs]{analog-to-digital convertors}
\acrodef{SNR}[SNR]{signal-to-noise ratio}
\acrodef{AWGN}[AWGN]{additive white Gaussian noise}
\acrodef{MRT}[MRT]{maximal ratio transmission}
\acrodef{DFT}[DFT]{Discrete Fourier Transform}
\acrodef{ULA}[ULA]{uniform linear array}
\acrodef{UPA}[UPA]{uniform planar array}
\acrodef{LS}[LS]{least squares}
\acrodef{ALMMSE}[ALMMSE]{approximate linear minimum mean squared error}
\acrodef{QIHT}[QIHT]{quantized iterative hard thresholding}
\acrodef{QIST}[QIST]{quantized iterative soft thresholding}
\acrodef{SVD}[SVD]{singular value decomposition}

\def\Frf{{\mathbf{F}_{\mathrm{RF}}}}
\def\Fbb{{\mathbf{F}_{\mathrm{BB}}}}
\def\Wrf{{\mathbf{W}_{\mathrm{RF}}}}
\def\rm#1{\mathrm{#1}}
\def\sf#1{\mathsf{#1}}
\begin{document}
\title{Bayesian Optimal Data Detector for Hybrid mmWave MIMO-OFDM Systems with Low-Resolution ADCs}
\author{Hengtao He,~\IEEEmembership{Student Member,~IEEE,}
       Chao-Kai Wen,~\IEEEmembership{Member,~IEEE,}
       and Shi Jin~\IEEEmembership{Senior Member,~IEEE}

\thanks{ Manuscript received September 15, 2017; revised January 25, 2018; accepted
March 04, 2018. Date of publication March
XX, 2018; date of current version May XX, 2018. The work of S. Jin was supported by the National Science
Foundation (NSFC) for Distinguished Young Scholars of China with Grant
61625106 and the NSFC with Grant 61531011. The work of C.-K. Wen was supported by the Ministry of Science
and Technology of Taiwan under Grants MOST 106-2221-E-110-019 and the ITRI in Hsinchu, Taiwan. The guest editor coordinating
the review of this paper and approving it for publication was Prof. L.
Dai. \emph{(Corresponding author: Shi Jin.)}}
\thanks{H. He and S. Jin are with the National Mobile Communications Research
Laboratory, Southeast University, Nanjing, 210096, China (e-mail: hehengtao@seu.edu.cn, and jinshi@seu.edu.cn).}
\thanks{C.-K.~Wen is with the Institute of Communications Engineering, National
Sun Yat-sen University, Kaohsiung 804, Taiwan (e-mail: chaokai.wen@mail.nsysu.edu.tw).}
}

\maketitle
\begin{abstract}
Hybrid analog-digital precoding architectures and low-resolution analog-to-digital converter (ADC) receivers are two solutions to reduce hardware cost and power consumption for millimeter wave (mmWave) multiple-input multiple-output (MIMO) communication systems with large antenna arrays.
 In this study, we consider a mmWave MIMO-OFDM receiver with a generalized hybrid architecture in which a small number of radio-frequency (RF) chains and low-resolution ADCs are employed simultaneously. Owing to the strong nonlinearity introduced by low-resolution ADCs, the task of data detection is challenging, particularly achieving a Bayesian optimal data detector. This study aims to fill
this gap. By using generalized expectation consistent signal recovery technique, we propose a computationally efficient data detection algorithm that provides a minimum mean-square error
 estimate on data symbols and is extended to a mixed-ADC architecture. Considering particular structure of MIMO-OFDM channel matirx, we provide a low-complexity realization in which only FFT operation and matrix-vector multiplications are required. Furthermore, we present an analytical framework to study the theoretical performance of the detector in the large-system limit, which can precisely evaluate the performance expressions such as mean-square error and symbol error rate. Based on this optimal detector, the potential of adding a few low-resolution RF chains and high-resolution ADCs for mixed-ADC architecture is investigated. Simulation results confirm the accuracy of our theoretical analysis and can be used for system design rapidly. The results reveal that adding a few low-resolution RF chains to original unquantized systems can obtain significant gains.



\end{abstract}

\section{Introduction}\label{sec_Introduction}
Millimeter wave spectrum in the range of $30$-$300$GHz enables the use of multi-gigahertz bandwidth that offers a higher data rates \cite{swindlehurst2014millimeter,bai2015coverage,RHeath2016JSTSP}, which have been recommended as an important part of
the $5$G mobile network. The arrays discussed for mmWave communication may be large due to the small wavelength, such as $16$ \cite{Cudak2014} or $256$  \cite{Rebeiz2015} elements.
 Leveraging the large antenna arrays employed at the transmitter and receiver, mmWave systems can perform directional beamforming to achieve high beamforming gain, which helps overcome large free-space pathloss of mmWave signals and guarantees sufficient received signal-to-noise ratio (SNR). 
 Unfortunately, the high hardware cost and power consumption are unaffordable when a dedicated RF chain is used for every antenna. Several massive MIMO architectures have been proposed to overcome this challenge. The first is the hybrid analog and digital precoding scheme, which uses various analog approaches such as phase shifters \cite{venkateswaran2010analog}, switches \cite{Wang2008swithes}, or lens \cite{zneg2016lens} to substantially reduce the number of RF chains. Another way is to use low-resolution ADCs at the receiver\cite{risi2014massive,fan2015uplink,Mo2015TSP,zhang2016spectral}.
\subsection{Related work}\label{Related work}
Hybrid architectures employ a smaller number of RF chains than the number of antennas to reduce power consumption and system complexity.
These architectures have been proposed for both low-frequency massive MIMO \cite{zhang2005variable,tan17} and mmWave systems \cite{el2014spatially,alkhateeb2014channel,Sohrabi2016Hybrid,Gao2016Hybrid}.
However, a common limitation of the hybrid architectures is the assumption that the receiver RF chains include high-resolution ADCs\footnote{Modern communication systems usually use $8$-$12$ bits ADCs to process received signals, for ease of analysis, we refer to high-resolution ADCs as infinite-resolution ADCs.}, which are power-hungry devices, especially when large bandwidth is involved. The power consumption of a typical ADC roughly scales linearly with the bandwidth and grows exponentially with the quantization bits \cite{lee2008analog}.

Thus, an alternative is to use low-resolution ADCs (e.g.,1-3 bits) to replace high-resolution ADCs, thereby resulting in quantized MIMO systems \cite{risi2014massive,fan2015uplink,Mo2015TSP,zhang2016spectral}. Only $\frac{2}{\pi}$ (1.96 dB) loss of mutual information is incurred in the low-SNR regime with 1-bit ADCs \cite{mezghani2007ultra}, which demonstrates its cost and energy efficiency. Furthermore, several studies have focused on capacity analysis \cite{fan2015uplink,Mo2015TSP,zhang2016spectral,Björnson2016twc,Fan2016Hardware}, energy efficiency\cite{Bai2015Energy,Orhan201515ITA,Verenzuela2016Hardware}, channel estimation \cite{risi2014massive,Mo14ACSSP}, and data detection \cite{Wang15ICC,CKWen2016TSP,Choi16TCOM} for quantized MIMO systems. However, strong nonlinear distortion caused by low-resolution ADCs inevitably causes problems such as high pilot overhead for channel estimation\cite{risi2014massive} and error floor for multi-user detection \cite{Wang15ICC,Choi16TCOM}.

Motivated by the aforementioned concerns, a mixed-ADC architecture for massive MIMO has been proposed \cite{liang2016mixed,liang2016TWC,Zhang2016TWC,Zhang2017mixed}, in which most antennas are installed with low-resolution ADCs while a few antennas are equipped with high-resolution ADCs. In \cite{liang2016mixed}, the generalized mutual information has been investigated to demonstrate
that the mixed-ADC architecture is able to achieve a large fraction of the channel capacity of ideal ADC architectures.
An optimal data detector for mixed-ADC massive MIMO is proposed in \cite{Zhang2016TWC} by using generalized approximate message passing (GAMP) algorithm  \cite{Rangan2012GAMP}, and two simple suboptimal detectors are also involved for complexity reduction. In \cite{Zhang2017mixed}, the spectral and energy efficiency of the mixed-ADC architecture are investigated, showing that
this type of architecture can achieve a better energy-rate tradeoff than the ideal high-resolution and pure low-resolution ADC architecture.
These studies confirm that the mixed-ADC architecture provides advantages for massive MIMO systems.

The architectures considered in the previous studies exhibit two extreme cases in which either a small number of RF chains with high-resolution ADCs, or the number of low-resolution RF chains equal to the number of antennas is assumed. Recently, \cite{mo2016hybrid} proposed a generalized architecture with hybrid beamforming and low-resolution ADCs, demonstrating that the achievable rate is comparable to that obtained by full-precision ADC receivers at low and medium SNRs. Moreover, an adaptive ADC design for a hybrid architecture is proposed, and the corresponding ADC bit allocation algorithm is derived to improve communication performance \cite{choi2017adaptive}. However, these studies only consider the spectral and energy efficiency of the generalized architecture.

\subsection{Contributions}\label{Contributions}
 In this study, we consider a hybrid architecture with low-resolution ADCs for a mmWave MIMO-OFDM receiver, and also investigate a mixed-ADC architecture. Owing to strong nonlinear distortion caused by coarse quantization on the received signals, the tasks of data detection is more challenging.
Several studies focus on data detection for quantized MIMO systems \cite{Wang15ICC,CKWen2016TSP,Choi16TCOM,Zhang2016TWC}, nevertheless, they only emphasize frequency-flat fading channels. For frequency-selective channels, \cite{Studer2016TCOM} develops channel estimation and data detection algorithms for quantized massive MIMO-OFDM based on mismatched quantization models, and in \cite{Mollen2017TWC}, uplink performance of wideband massive MIMO with one-bit ADC is analyzed.
 These studies only focus on low-frequency massive MIMO systems. In fact, in a low-frequency massive MIMO system, an increasing number of receiving antennas is robust against coarse quantization, using a complex nonlinear detector in the receiver is unnecessary because a traditional linear detector, such as linear MMSE (LMMSE) detector is competent. However, in mmWave MIMO systems, limited RF chains are in the receiver, which means that observations are limited. Under the circumstances, achieving optimal detection is significantly important.

In \cite{wang2017Gturbo}, an optimal and computationally tractable data detector based on turbo iteration principle is proposed for the mmWave quantized SISO-OFDM system. However, higher data rates are needed for mmWave wireless systems, which mainly consider MIMO communication with multi-stream and multi-user scenarios. Furthermore, an optimal detector is necessary for performance analysis and design optimization. To the best of our knowledge, no study has been conducted to exploit the optimal detector for mmWave quantized MIMO-OFDM systems.
The present study takes the first step toward this direction. Our work can also be interpreted as a solution to the detection problem in single-carrier frequency division multiple access systems with coarse quantization relevant to the 3GPP LTE uplink \cite{3GPPLTE} which is provided as a future study in \cite{Studer2016TCOM}. Nevertheless, we focus on the systems in which RF chains in the transmitter and receiver are comparable, which are significantly different from the massive MIMO systems presented in \cite{Studer2016TCOM}.

    In this article, we propose a Bayesian optimal data detector for mmWave MIMO-OFDM systems with low-resolution ADCs; this detector is based on generalized expectation consistent signal recovery (GEC-SR) algorithm in our previous conference work \cite{He2017GEC}, and is extended to a mixed-ADC architecture. Considering this optimal detector, we determine how much gain can obtain from adding a few low-resolution RF chains, and demonstrate how many high-resolution ADCs are needed to render the quantized distortion acceptable when the transmitter and receiver have a comparable number of RF chains. The main contributions of this study are summarized as follows.

\begin{itemize}
  \item Achieving a Bayes-optimal data detector for mmWave quantized MIMO-OFDM system is very challenging. We apply a computationally tractable GEC-SR algorithm for this problem, which is easily extended to a mixed-ADC architecture. By exploiting a particular structure of the sensing matrix, we provide a low-complexity realization to reduce computational complexity.
  \item We develop an analytical framework to analyze the performance of the GEC-SR algorithm, which is consistent with the theoretical Bayes-optimal estimator obtained by replica method from statistical physics \cite{Nishimori2001Statistical}. This consistency indicates the optimality of the proposed detector. Simulation results verify the accuracy of our analysis. With theoretical analysis, performance metrics, such as the average MSE and SER, can be analytically determined without time-consuming Monte Carlo simulation.
  \item Simulation results show that the proposed detector outperforms the representative GAMP-based detectors in quantized MIMO-OFDM systems.
  Based on this optimal detector, we investigate the potential of adding a few low-resolution RF chains and high-resolution ADCs for mixed-ADC architecture. We obtain several useful observations on the basis of our analysis for system design. We demonstrate that in unquantized MIMO-OFDM systems, adding a few low-resolution RF chains can obtain significant gains, especially adding $1$-bit RF chains, which are easily implemented for hardware. We also show that adopting a few high-resolution ADCs in quantized MIMO-OFDM systems can bring remarkable gains, and error floor due to nonlinear distortion is eliminated.
 \end{itemize}
\begin{figure*}
  \centering
  \includegraphics[width=16cm]{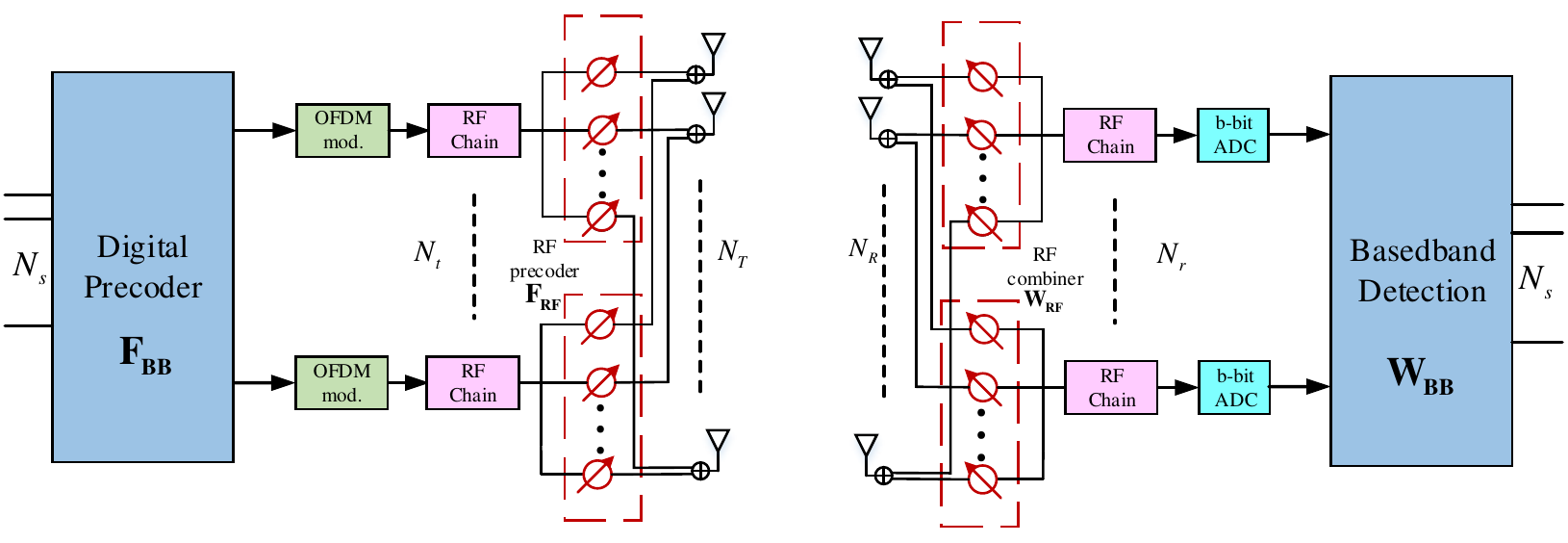}
  \caption{.~~Hybrid precoding and combining architecture in a MIMO-OFDM system with low-resolution ADCs.}\label{system model}
\end{figure*}

\emph{Notations}---For any matrix $\mathbf{A}$, $\mathbf{A}^{H}$ is the conjugate transpose of $\mathbf{A}$,
and $\mathsf{tr}(\mathbf{A})$ denotes the traces of $\mathbf{A}$. In addition, $\mathbf{0}$ is the zero matrix, $\mathbf{1}$ is all-ones vector, $\mathrm{Diag}(\mathbf{v})$ is the diagonal matrix whose diagonal equals $\mathbf{v}$, and $\mathbf{d}(\mathbf{Q})$ denotes the diagonalization operator
that returns vector $\bd(\bQ):= \frac{\mathsf{tr}(\bQ)}{N}\mathbf{1}$ for $\bQ\in\bbC^{N\times N}$.
In addition, $\otimes$ denotes the Kronecker product, and $\oslash$ and $\odot$ denote componentwise vector division and vector multiplication, respectively.
A random vector $\mathbf{z}$ drawn from the proper complex Gaussian distribution of mean $\boldsymbol{\mu}$ and covariance $\boldsymbol{\Omega}$ is described by the following probability density function:
\begin{equation*}
  \mathcal{N}_{\mathbb{C}}(\mathbf{z};\boldsymbol{\mu},\boldsymbol{\Omega})=\frac{1}{\mathrm{det}(\pi \boldsymbol{\Omega})}
  e^{-(\mathbf{z}-\boldsymbol{\mu})^{H}\boldsymbol{\Omega}^{-1}(\mathbf{z}-\boldsymbol{\mu})}.
\end{equation*}
We use $\rmD z$ to denote the real Gaussian integration measure
\begin{equation*}
  Dz=\phi(z)dz, \quad \phi(z)\triangleq\frac{1}{\sqrt{2\pi}}e^{-\frac{z^{2}}{2}},
\end{equation*}
and we use $\rmD z_{c}=\frac{e^{-|z|^{2}}}{\pi}dz$ to denote the complex Gaussian integration measure. Finally, $\Phi(x)\triangleq \int_{-\infty}^{x} \rmD z$
denotes the cumulative Gaussian distribution function.

\section{System Model}\label{System Model}
In this paper, we consider a hybrid precoding architecture in a MIMO-OFDM system with low-resolution ADCs, as shown in Fig. \ref{system model}.
The transmitter and receiver are equipped with $N_{\mathrm{T}}$ and $N_{\mathrm{R}}$ antennas, respectively. The transmitter is assumed to have $\Nct$  RF chains with full-precision digital-to-analog converters (DACs), while the receiver employs $\Ncr$ RF chains with low-resolution (1-3 bits) ADCs. Furthermore, the number of antennas and RF chains are assumed to satisfy $\left( \Nct \leq \Nt, \Ncr \leq \Nr \right)$. The transmitter and receiver communicate through $\Ns$ data streams, with $\Ns = \Nct$, and $\Ncr \geq \Nct $. Meanwhile, we consider the OFDM system with $N_{c}$ orthogonal subcarriers.

\textbf{Remark 1}: Precoding techniques operating in the RF domain will be used at the transmitter and receiver to overcome the high propagation loss in the mmWave band \cite{wang2017Gturbo}. We can use an equivalent representation for the input-output relationship of the hybrid mmWave MIMO architecture using analog precoding and receiver combining. Specifically, an equivalent MIMO channel between transmitter and receiver RF chains is expressed by
\begin{equation}\label{eqivaMIMO}
\tilde{\bH}= \Wrf\bG\Frf
\end{equation}
where $\Wrf\in\bbC^{N_{r}\times N_{R}}$ and $\Frf\in\bbC^{N_{T}\times N_{t}}$ are the analog combiner and precoder, respectively; and $\bG\in\bbC^{N_{R}\times N_{T}}$ represents the channel response matrix between all transmit and receive antennas. Therefore, based on OFDM technology, an equivalent quantized MIMO-OFDM system is established, which is composed of
 $\Nc$ subcarriers, $\Ncr$ receive antennas, $\Nct$ transmit antennas, and low-resolution ADCs at the receiver. Then, we will formulate our problems based on this equivalent system model.\hfill\ensuremath{\blacksquare}

   We consider a quantized MIMO-OFDM system with $N_t$ transmit antennas, $N_r$ receive antennas and $N_c$ subcarriers. 
   At the transmitter of the MIMO-OFDM system, the coded bits are first mapped into the complex symbol sequence $\mathbf{x}\in\mathbb{C}^{N_t N_c\times 1}$. To reduce the peak-to-average power ratio of the transmitted signals, we adopt simple discrete Fourier
   transform (DFT) precoding in the digital domain. Therefore, the equivalent complex symbol sequence is denoted by
   \begin{equation}\label{eqFBBx}
      \bar{\mathbf{x}}=\mathbf{F}_{\mathbf{BB}} \mathbf{x},
   \end{equation}
   where $\mathbf{F}_{\mathbf{BB}}\in\mathbb{C}^{N_t N_c\times N_t N_c}$ is a unitary DFT matrix with rows that are randomly permutated. The equivalent complex symbol sequence $ \bar{\mathbf{x}}$ is then divided into $N_t$ streams $\bar{\mathbf{x}}_{n_t}$. After the application of an IFFT operation, the frequency-domain signal $\bar{\mathbf{x}}_{n_t}$ is transformed to the time domain. The transmitted signal is filtered by a multipath channel, which can be represented by a tapped delay line model with $L$ taps for each pair of transmit antenna $n_{t}\in \{1,\ldots N_t\}$ and receive antenna $n_{r}\in \{1,\ldots N_r\}$, denoted by $\mathbf{h}_{n_{r}n_{t}}\in\mathbb{C}^{L}$. with the use of a cyclic prefix (CP), each pair of antennas has an equivalent circulant channel convolution matrix $\mathbf{H}_{n_{r}n_{t}}\in\mathbb{C}^{N_c\times N_c}$ whose first column is
$[\mathbf{h}_{n_{r}n_{t}}^{T}, \mathbf{0}^{T}_{(N_c-L)\times 1}]^{T}$.

At the receiver, the unquantized received block of the OFDM symbol in each receive antenna can be written as
\begin{equation}\label{OFDM}
  \by_{nr}=\sum\limits_{n_{t}=1}^{\Nct}\bH_{n_{r}n_{t}}\bF^{H}\bar{\bx}_{n_{t}}+\bn_{n_{r}},
\end{equation}
where $\bF \in\bbC^{N_c\times N_c}$ denotes a unitary DFT matrix in which $(k,l)$th entry is $\frac{1}{\sqrt{\Nc}}e^{-2\pi j(k-1)(l-1)/\Nc}$, and $\bn_{n_{r}}\sim\cN_{\bbC}(0,\sigma_{N}^{2}\mathbf{I})$ is the white Gaussian noise.
The quantized received signal is denoted as
\begin{equation}\label{Quantized_OFDM}
  \tilde{\by}_{n_{r}}=\cQ_{c}\left(\sum\limits_{n_{t}=1}^{\Nct}\bH_{n_{r}n_{t}}\bF^{H}\bar{\bx}_{n_{t}}+\bn_{n_{r}} \right),
\end{equation}
where $\cQ_{c}$ is the complex-valued quantizer, and the circulant channel convolution matrices $\bH_{n_{r}n_{t}}$  can be decomposed as
\begin{equation}\label{convolution_matrices}
\bH_{n_{r}n_{t}}=\bF^{H}\mathbf{\Lambda}_{n_{r}n_{t}}\bF,
\end{equation}
where $\mathbf{\Lambda}_{n_{r}n_{t}}$ is a diagonal matrix with diagonal elements that are frequency-domain responses. Meanwhile, we can rewrite (\ref{Quantized_OFDM}) as
\begin{equation}\label{OFDM2}
  \tilde{\by}_{n_{r}}=\cQ_{c}\left(\sum\limits_{n_{t}=1}^{\Nct}\bF^{H}\mathbf{\Lambda}_{n_{r}n_{t}}\bar{\bx}_{n_{t}}+\bn_{n_{r}} \right),
\end{equation}

For ease of analysis, we can reformulate the model in (\ref{OFDM2}) using
the multiple measurement vector (MMV) framework \cite{Prasad2015TSP}. By using simple matrix arrangement, the final MMV framework is given by
\begin{equation}\label{system_model}
  \tilde{\by}= \cQ_{c}(\mathbf{A}\bx+\bn),
\end{equation}
where the channel matrix is denoted by
\begin{equation}\label{channel_matrix}
\bA=\tilde{\bA}\Fbb=\left(\begin{array}{c c c}
\mathbf{F}^H\mathbf{\Lambda}_{11} & \cdots & \mathbf{F}^H\mathbf{\Lambda}_{1\Nct} \\
\vdots & \ddots & \vdots \\
\mathbf{F}^H\mathbf{\Lambda}_{\Ncr1} & \cdots & \mathbf{F}^H\mathbf{\Lambda}_{\Ncr \Nct}
\end{array}\right)\Fbb.
\end{equation}
And the observation vector can be expressed as $\tilde{\by}=[\tilde{\by}_{1}^{T},\ldots,\tilde{\by}_{\Ncr}^{T}]^{T}$.
The noise $\bn$ is defined in the same way as $\tilde{\by}$. The dimensions of $\tilde{\by}$, $\bx$, and $\bA$ are $\Nc\Ncr\times1$, $\Nc\Nct\times1$, and $\Nc\Ncr\times\Nc\Nct$, respectively. In addition, for ease of notation, we define
\begin{equation}\label{MN}
  M=\Nc\Ncr,~~\mathrm{and}~~N=\Nc\Nct.
\end{equation}
This equivalent quantized MIMO-OFDM system is useful to formulate data detection problem for hybrid broadband mmWave systems. An optimal data detection algorithm will be introduced in the next section.

In the current paper, we mainly focus on  a typical uniform midrise quantizer with quantization step size $\Delta$. Such a quantizer maps a real-valued\footnote{For ease of
notation, we abuse $\tilde{y}$ to denote each real channel although it should be specified as ${\mathrm{Re}}(\tilde{y})$ or ${\mathrm{Im}}(\tilde{y})$, and omit index $n$.} input that falls in interval $(\tilde{y}-\frac{\Delta}{2},
\tilde{y}+\frac{\Delta}{2}]$ to value $\tilde{y}$ from the discrete set
\begin{equation} \label{2.2}
    \cR_{{\kappa}} \triangleq {\left\{ \Big({-\frac{1}{2}}+b\Big) \Delta; \,\, b=-\frac{2^{{\kappa}}}{2}+1, \cdots, \frac{2^{{\kappa}}}{2} \right\}} ,
\end{equation}
where $\kappa$ is the quantization bits.

For notational convenience, we simply express the lower and upper thresholds associated with $\tilde{y}$ as $r^{\mathrm{low}}$ and $r^{\mathrm{up}}$, respectively; specifically, they are
\begin{subequations}\label{quanbound}
    \begin{equation}
    r^{\mathrm{low}} = \left\{ \begin{array}{ll}
    \tilde{y}-\frac{\Delta}{2}, & \textrm{for $\tilde{y}\ge -{\left(\frac{2^{\kappa}}{2}-1\right)}\Delta$},\\
    -\infty, & \textrm{otherwise},
    \end{array} \right.
    \end{equation}
    and
    \begin{equation}
    r^{\mathrm{up}} = \left\{ \begin{array}{ll}
    \tilde{y}+\frac{\Delta}{2}, & \textrm{for $\tilde{y}\le {\left(\frac{2^{\kappa}}{2}-1\right)}\Delta$},\\
    \infty, & \textrm{otherwise}.
    \end{array} \right.
    \end{equation}
\end{subequations}



\section{Bayesian Data Detector}\label{Bayesian Data Detector}
In this section, we introduce a theoretical problem description of quantized MIMO-OFDM detection with perfect CSI. We analyze this detection problem under the framework of Bayesian inference, which provides a foundation to achieve the best MSE estimates. However, the computational complexity of Bayesian inference is extremely high. Fortunately, the GEC-SR algorithm presents a computationally efficient way to achieve Bayes-optimal performance.

\subsection{Bayesian Detector}\label{Bayesian Detector}
We adopt the Bayesian inference to recover the signals $\bX$ from the quantized measurements $\tilde{\bY}$. Based on Bayes theorem, the posterior probability is given by
\begin{equation}\label{eq_Bayesian}
  \sfP(\bX|\tilde{\bY},\bA)=\frac{\sfP_{\mathrm{out}}(\tilde{\bY}|\bX,\bA)\sfP(\bX)}{\sfP(\tilde{\bY}|\bA)},
\end{equation}
where $\sfP_{\mathrm{out}}(\tilde{\bY}|\bX,\bA)$ is the likelihood function, $\sfP(\bX)$ is the prior distribution of the data, and $\sfP(\tilde{\bY}|\bA)=\int\sfP_{\mathrm{out}}(\tilde{\bY}|\bX,\bA)\sfP(\bX)\mathrm{d}\bX$ is the marginal distribution. We consider perfect CSI in the receiver, and separable prior and likelihood function. Notably, the likelihood function is given by
\begin{equation}\label{eq_likelihood}
\sfP_{\mathrm{out}}(\tilde{\bY}|\bX,\bA)=\prod\limits_{j=1}^{M}\sfP_{\mathrm{out}}(\tilde{Y}_{j}|Z_{j}),
\end{equation}
where $Z_j$ is the $j$-th entry of the noiseless observation vector $\mathbf{Z}$ with $\mathbf{Z}=\mathbf{A}\mathbf{X}$.
According to the property of the complex-valued quantizer, we obtain
\begin{equation}\label{eq_likelihood_separable}
  \sfP_{\mathrm{out}}(\tilde{Y}_{j}|Z_{j})=\Psi_{b}(\mathrm{Re}(Z_{j}))\Psi_{b'}(\mathrm{Im}(Z_{j})),
\end{equation}
where
\begin{equation}\label{PHIfunction}
\Psi_{b}(x)\triangleq\Phi(\frac{\sqrt{2}(r^{\mathrm{up}}-x)}{\sigma_{N}})-\Phi(\frac{\sqrt{2}(r^{\mathrm{low}}-x)}{\sigma_{N}}).
\end{equation}
Given the posterior probability $\sfP(\bX|\tilde{\bY},\bA)$, the Bayesian (MMSE) estimate is obtained by
\begin{equation}\label{MMSE_estimate}
\hat{\bX}=\int\bX \sfP(\bX|\tilde{\bY},\bA)d\bX,
\end{equation}
and its $j$-th element is expressed by
\begin{equation}\label{MMSE_element}
  \hat{X}_{j}=\sfE\{X_{j}|\tilde{\bY},\bA\},
\end{equation}
where the expectation over $X_{j}$ is w.r.t. the marginal posterior probability
\begin{equation}\label{MMSE_marginal}
  \sfP(X_{j}|\tilde{\bY},\bA)=\int_{\bX\setminus X_{j}}\sfP(\bX|\tilde{\bY},\bA)d\bX.
\end{equation}

However, the Bayesian MMSE estimator is not computationally tractable because the marginal posterior probability in (\ref{MMSE_marginal}) involves a high-dimensional integral. In our recent study \cite{He2017GEC}, the innovative generalized expectation consistent signal recovery (GEC-SR) algorithm was proposed as an iterative method to recover signal $\bx$ from nonlinear measurements $\tilde{\by}$.
We will show the GEC-SR algorithm for quantized MIMO-OFDM systems in the following subsection.
\begin{figure*}[t]
  \centering
  \includegraphics[width=14cm]{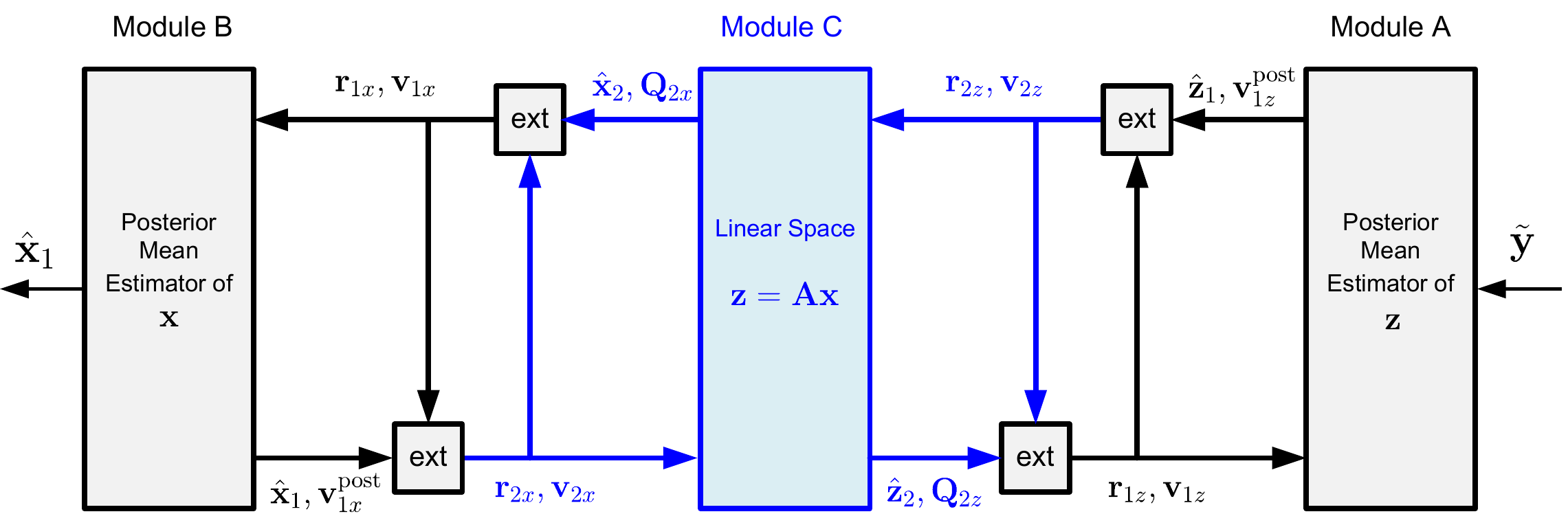}
  \caption{.~~Block diagram of GEC-SR algorithm.}\label{GEC-diagram}
\end{figure*}
\subsection{GEC-based Algorithm}\label{GEC-based Algorithm}

We present the GEC-based data detection algorithm for quantized MIMO-OFDM system in Algorithm 1. The block diagram is illustrated in Fig. \ref{GEC-diagram}, which consists of three modules: A, B and C. Module A computes the posterior mean and variance of $\mathbf{z}$, which is interpreted as \emph{de-quantization} process, module B computes the posterior mean and variance of $\mathbf{x}$ by considering the prior distribution, and module C provides the framework that constrains the estimation problem into the linear space $\mathbf{z} = \mathbf{A}\mathbf{x}$.

\begin{algorithm}\label{algGE}\small{
\caption{GEC-SR for the quantized MIMO-OFDM} 
\hspace*{0.02in} {\bf Input:} 
Quantized measurements $\tilde{\mathbf{y}}$, sensing matrix $\mathbf{A}$, likelihood $\sfP(\tilde{\mathbf{y}}|\mathbf{z})$, and prior distribution $\sfP(\mathbf{x})$.\\
\hspace*{0.02in} {\bf Output:} 
Recovered signal $\hat{\mathbf{x}}_{1}$.\\
\hspace*{0.02in} {\bf Initialize:}
$t \leftarrow 1$, $\mathbf{r}_{1\mathbf{z}}\leftarrow \mathbf{0}$, $\mathbf{r}_{2\mathbf{x}}\leftarrow \mathbf{0}$, $\mathbf{v}_{1\mathbf{z}}\leftarrow P_{z} \mathbf{1}$, and $\mathbf{v}_{2\mathbf{x}}\leftarrow P_{x} \mathbf{1}$.

 \For{$t = 1,\cdots, t_{\max}$ }{

 \textbf{Module A:}

(1) Compute the posterior mean and covariance of $\mathbf{z}$\\
\nl     $\hat{ \mathbf{z}}_{1} =\mathtt{E}\left\{\mathbf{z}|\mathbf{\mathbf{r}}_{1\mathbf{z}},\mathbf{v}_{1\mathbf{z}} \right\},$\\
\nl    $\mathbf{v}_{1\mathbf{z}}^{\mathrm{post}} = \mathtt{Var} \left\{ \mathbf{z}|\mathbf{\mathbf{r}}_{1\mathbf{z}},\mathbf{v}_{1\mathbf{z}} \right\}.$

(2) Compute the extrinsic information of $\mathbf{z}$\\
 \nl$\mathbf{v}_{2\mathbf{z}}= \mathbf{1}\oslash {\left( \mathbf{1}\oslash\mathbf{v}_{1\mathbf{z}}^{ \mathrm{post}}- \mathbf{1}\oslash\mathbf{v}_{1\mathbf{z}} \right)}, $\\
 \nl$\mathbf{r}_{2\mathbf{z}}= \mathbf{v}_{2\mathbf{z}} \odot {\left( \hat{\mathbf{z}}_{1}\oslash\mathbf{v}_{1\mathbf{z}}^{ \mathrm{post}}-\mathbf{r}_{1\mathbf{z}}\oslash\mathbf{v}_{1\mathbf{z}} \right)}.$

   \textbf{Module C:}

(3) Compute the mean and covariance of $\mathbf{x}$  from the linear space\\
   \nl  $\mathbf{Q}_{2\mathbf{x}}={\left(\mathrm{Diag}(\mathbf{1}\oslash \mathbf{v}_{2\mathbf{x}})+\mathbf{A}^{H}\mathrm{Diag}(\mathbf{1}\oslash\mathbf{v}_{2\mathbf{z}})\mathbf{A} \right)}^{-1}, $ \\
  \nl $\hat{\mathbf{x}}_{2}=\mathbf{Q}_{2\mathbf{x}} \left(\mathbf{r}_{2\mathbf{x}}\oslash\mathbf{v}_{2\mathbf{x}}+\mathbf{A}^{H}\mathbf{r}_{2\mathbf{z}}\oslash\mathbf{v}_{2\mathbf{z}}\right).$

(4) Compute the extrinsic information of $\mathbf{x}$\\
  \nl   $\mathbf{v}_{1\mathbf{x}}=\mathbf{1}\oslash \left( \mathbf{1}\oslash\mathbf{d}(\mathbf{Q}_{2\mathbf{x}}) - \mathbf{1}\oslash\mathbf{v}_{2\mathbf{x}} \right),$\\
  \nl   $\mathbf{r}_{1\mathbf{x}}=\mathbf{v}_{1\mathbf{x}} \odot \left( \hat{\mathbf{x}}_{2}\oslash\mathbf{d}(\mathbf{Q}_{2\mathbf{x}}) - \mathbf{r}_{2\mathbf{x}}\oslash\mathbf{v}_{2\mathbf{x}} \right).$

 \textbf{Module B:}

(5) Compute the mean and covariance of $\mathbf{x}$ \\
   \nl   $\hat{\mathbf{x}}_{1}=\mathtt{E}\left\{\mathbf{x}|\mathbf{\mathbf{r}}_{1\mathbf{x}},\mathbf{v}_{1\mathbf{x}}\right\}, $\\
   \nl   $\mathbf{v}_{1\mathbf{x}}^{\mathrm{post}}=\mathtt{Var} \left\{ \mathbf{x}|\mathbf{\mathbf{r}}_{1\mathbf{x}},\mathbf{v}_{1\mathbf{x}} \right\}. $

(6) Compute the extrinsic information of $\mathbf{x}$\\
 \nl    $\mathbf{v}_{2\mathbf{x}}= \mathbf{1}\oslash {\left( \mathbf{1}\oslash\mathbf{v}_{1\mathbf{x}}^{\mathrm{post}}- \mathbf{1}\oslash\mathbf{v}_{1\mathbf{x}} \right)}, $\\
  \nl   $\mathbf{r}_{2\mathbf{x}}=\mathbf{v}_{2\mathbf{x}} \odot {\left( \hat{\mathbf{x}}_{1}\oslash\mathbf{v}_{1\mathbf{x}}^{\mathrm{post}} - \mathbf{r}_{1\mathbf{x}}\oslash\mathbf{v}_{1\mathbf{x}} \right)} . $

 \textbf{Module C:}

(7) Compute the mean and covariance of $\mathbf{z}$ from the linear space\\
 \nl   $\mathbf{Q}_{2\mathbf{x}}={\left(\mathrm{Diag}(\mathbf{1}\oslash \mathbf{v}_{2\mathbf{x}})+\mathbf{A}^{H}\mathrm{Diag}(\mathbf{1}\oslash\mathbf{v}_{2\mathbf{z}})\mathbf{A} \right)}^{-1},$\\
 \nl   $ \hat{\mathbf{x}}_{2}=\mathbf{Q}_{2\mathbf{x}} \left(\mathbf{r}_{2\mathbf{x}}\oslash\mathbf{v}_{2\mathbf{x}}+\mathbf{A}^{H}\mathbf{r}_{2\mathbf{z}}\oslash\mathbf{v}_{2\mathbf{z}}\right),$\\
 \nl   $\mathbf{Q}_{2\mathbf{z}}=\mathbf{A}\mathbf{Q}_{2\mathbf{x}}\mathbf{A}^{H}, $\\
 \nl   $\hat{\mathbf{z}}_{2}=\mathbf{A}\hat{\mathbf{x}}_{2}. $

(8) Compute the extrinsic information of $\mathbf{z}$ \\
\nl  $\mathbf{v}_{1\mathbf{z}}= \mathbf{1}\oslash {\left(\mathbf{1}\oslash\mathbf{d}(\mathbf{Q}_{2\mathbf{z}}) - \mathbf{1}\oslash\mathbf{v}_{2\mathbf{z}} \right)},$ \\
\nl     $\mathbf{r}_{1\mathbf{z}}=\mathbf{v}_{1\mathbf{z}} \odot {\left( \hat{\mathbf{z}}_{2}\oslash\mathbf{d}(\mathbf{Q}_{2\mathbf{z}}) - \mathbf{r}_{2\mathbf{z}} \oslash \mathbf{v}_{2\mathbf{z}} \right)}. $
     }
     }
\end{algorithm}

These procedures follow a circular manner, that is, $A \rightarrow C \rightarrow B \rightarrow C \rightarrow A \rightarrow \cdots$.
In addition, each module uses the turbo principle in iterative decoding, that is, each module passes the extrinsic messages to its next module. The three modules are executed iteratively until convergence.

Before introducing the GEC-SR algorithm, we define two auxiliary variables:
\begin{equation}\label{eq:defPxPz}
  P_{x} = \mathsf{E}\{|x_{n}|^{2}\} ~~\mbox{and}~~  P_{z} = P_{x} \cdot \mathsf{tr}(\mathbf{A}^H\mathbf{A})/M,
\end{equation}
which are interpreted as the power of $x_{n}$ and $z_{n}$, respectively.

To better understand the algorithm, we provide detailed explanations for Algorithm 1. Lines 1--2 compute the posterior mean and variance of $z_{n}$ from quantized measurements $\tilde{y}_{n}$, and the expectation w.r.t. the posterior
\begin{equation}\label{posterior_estimate_z}
  \sfP_{Z}(z_{n}|\tilde{y}_{n})=\frac{\sfP_{\mathrm{out}}(\tilde{y}_{n}|z_{n})\sfP_{Z}(z_{n})}{\int\sfP_{\mathrm{out}}(\tilde{y}_{n}|z_{n})\sfP_{Z}(z_{n})dz_{n}},
\end{equation}
where $\sfP_{Z}(z_{n})$ is assumed to be $\cN_{\bbC}(z_{n};r_{1z,n},v_{1z,n})$, and the explicit expression that can be obtained by derivation of \cite[Appendix A]{CKWen2016TSP} is given by
\begin{align}
    \hat{z}_{1}
   &= r_{1z} + \frac{\sign(\tilde{y}) v_{1z} }{\sqrt{2(\sigma_{N}^2 + v_{1z})}} \left( \frac{\phi(\eta_1)-\phi(\eta_2)}{\Phi(\eta_1)-\Phi(\eta_2)} \right),
   \label{eq:hatZ_RealGaussian} \\
    v_{1z}^{\mathrm{post}} & = \frac{v_{1z}}{2} - \frac{(v_{1z})^2}{2(\sigma_{N}^2 + v_{1z})}\times  \nonumber \\
     & \left( \frac{\eta_1\phi(\eta_1)-\eta_2\phi(\eta_2)}{\Phi(\eta_1)-\Phi(\eta_2)}
     + \left(\frac{\phi(\eta_1)-\phi(\eta_2)}{\Phi(\eta_1)-\Phi(\eta_2)}\right)^2 \right),
     \label{eq:mseZ_RealGaussian}
\end{align}
where
\begin{subequations} \label{eq:eta_def}
\begin{align}
    \eta_1 &= \frac{\sign(\tilde{y})r_{1z}-\min\{|r^{\mathrm{low}}|,|r^{\mathrm{up}}|\}}{\sqrt{(\sigma_{N}^2 + v_{1z})/2}}, \\
    \eta_2 &= \frac{\sign(\tilde{y})r_{1z}-\max\{|r^{\mathrm{low}}|,|r^{\mathrm{up}}|\}}{\sqrt{(\sigma_{N}^2 + v_{1z})/2}}.
\end{align}
\end{subequations}
The real and imaginary parts are quantized separately, and each complex-valued channel can be decoupled into two real-valued channels. The expressions (\ref{eq:hatZ_RealGaussian}) and (\ref{eq:mseZ_RealGaussian}) are the estimators only for the real part of $\hat{z}_{1}$. To facilitate notation, we have
abused $\tilde{y}$ and $\hat{z}_{1}$ in (\ref{eq:hatZ_RealGaussian}) and (\ref{eq:mseZ_RealGaussian}) to denote $ \mathrm{Re}(\tilde{y})$ and $ \mathrm{Re}(\hat{z}_{1})$, respectively, and we omit index $n$ in the aforementioned expression. The estimator for the imaginary part $\mathrm{Im}(\hat{z}_{1})$ can be obtained analogously as (\ref{eq:hatZ_RealGaussian}) and (\ref{eq:mseZ_RealGaussian}), while $\tilde{y}$ and $b$ should be replaced by
$\mathrm{Im}(\tilde{y})$ and $b'$, respectively.
Lines $3$--$4$
compute the extrinsic information of $\bz$ using the turbo principle. Lines $5$--$6$ perform the LMMSE estimate of $\bx$ under the following assumption:
\begin{equation}\label{LMMSE_estimate}
  \br_{2\bz}=\bz_{2}+\bw_{2\bz},
\end{equation}
where $\bw_{2\bz} \sim \cN_{\bbC}(\mathbf{0},\mathrm{Diag}(\bv_{2\bz}))$, $\bz_{2}=\bA\bx_{2}$, and $\bx_{2} \sim \cN_{\bbC}(\bx_{2};\br_{2\bx},\mathrm{Diag}(\bv_{2\bx}))$. Lines $7$--$8$ compute the extrinsic information of $\bx$ and pass it to module B as a prior information.
Lines $9$--$10$ estimate the posterior mean $\hat{\bx}_{1}$ and variance $\bv_{1\bx}^{\mathrm{post}}$ by considering the true prior $\sfP(\mathbf{x})$, which is assumed to estimate $\bx$ from an AWGN observation that is,
\begin{equation}\label{AWGN_estimate}
\br_{1\bx}=\bx+\bw_{1\bx},
\end{equation}
where $\bw_{1\bx} \sim \cN_{\bbC}(\mathbf{0},\mathrm{Diag}(\bv_{1\bx}))$, and lines $11$--$12$ compute the extrinsic information of $\bx$ using the turbo principle. Lines $13$--$16$ constrain the estimated problem into a linear space $\bz=\bA\bx$ which performs the same procedure as Lines $5$--$6$. Lines $17$--$18$ compute the extrinsic information of $\bz$ and passes it to module A as prior information.

\subsection{OFDM-based Low-complexity Realization}\label{SVD-Form}
Although three modules are involved in Algorithm 1, most of lines can be computed by element-wise, except for
lines $5$ and $13$ which need matrix inverse to calculate the covariance matrix $\bQ_{2\bx}$. In fact, only diagonal elements of the covariance matrix are required to be passed to the next module, which motivates us to provide a low-complexity realization by exploiting a particular
structure of the sensing matrix.
Notably, the equivalent channel matrix is defined by
\begin{align}\label{channel_matrix}
\tilde{\bA}&=\left(\begin{array}{c c c}
\mathbf{F}^H\mathbf{\Lambda}_{11} & \cdots & \mathbf{F}^H\mathbf{\Lambda}_{1\Nct} \\
\vdots & \ddots & \vdots \\
\mathbf{F}^H\mathbf{\Lambda}_{\Ncr1} & \cdots & \mathbf{F}^H\mathbf{\Lambda}_{\Ncr \Nct}
\end{array}\right) \nonumber \\
&=(\bI\otimes\mathbf{F}^H)\underbrace{\left(\begin{array}{c c c}
\mathbf{\Lambda}_{11} & \cdots & \mathbf{\Lambda}_{1\Nct} \\
\vdots & \ddots & \vdots \\
\mathbf{\Lambda}_{\Ncr1} & \cdots & \mathbf{\Lambda}_{\Ncr \Nct}
\end{array}\right)}_{\triangleq{\bC}},
\end{align}
%
where matrix $\bC$ contains all frequency-domain channel responses between each pair of antennas. $\mathbf{\Lambda}_{n_{r}n_{t}}$ is a diagonal matrix and its elements are denoted by $[\lambda_{n_{r}n_{t}}^{1}, \ldots, \lambda_{n_{r}n_{t}}^{N_{c}}]$. After elementary transformation of matrix $\bC$, we obtain a block diagonal matrix $\bD$, which is given by
\begin{align}\label{blockdiagmtx}\small
  \bD & \triangleq\bP \bC \bQ= \nonumber  \\
  & \left(\begin{array}{c c c}
\begin{array}{c c c}
\lambda_{11}^{1} & \cdots & \lambda_{1\Nct}^{1} \\
\vdots & \ddots & \vdots \\
\lambda_{\Ncr1}^{1} & \cdots & \lambda_{\Ncr \Nct}^{1}
\end{array} & \cdots & \mathbf{0} \\
\vdots & \ddots & \vdots \\
\mathbf{0} & \cdots &
\begin{array}{c c c}
\lambda_{11}^{\Nc} & \cdots & \lambda_{1\Nct}^{\Nc} \\
\vdots & \ddots & \vdots \\
\lambda_{\Ncr1}^{\Nc} & \cdots & \lambda_{\Ncr \Nct}^{\Nc}
\end{array}
\end{array}\right),
\end{align}
where $\bP$ and $\bQ$ are permutation matrices that are known at the receiver once the channel is known. Each block $\bD_{i}$ contains a frequency response of all pairs of antennas in a corresponding subcarrier. We obtain the SVD of each block as follows,
\begin{equation}\label{blocksvd}
\bD_{i}=\bU_{i}\mathbf{\Sigma}_{i}\bV_{i}^{H}
\end{equation}
Therefore, the final expression of $\bA$ is given by
\begin{align}\label{finalSVD}
  \bA&=\underbrace{(\bI\otimes\mathbf{F}^H)\bP^{-1}\left[\begin{array}{c c c}
\bU_{1}  \\
 &\ddots   \\
 &&\bU_{\Nc}
\end{array}\right]}_{\triangleq{\bU}}\times  \nonumber  \\
&\underbrace{\left[\begin{array}{c c c}
\mathbf{\Sigma}_{1}  \\
 &\ddots   \\
 &&\mathbf{\Sigma}_{\Nc}
\end{array}\right]}_{\triangleq{\mathbf{\bS}}}
\underbrace{\underbrace{\left[\begin{array}{c c c}
\bV_{1}^{H}  \\
 &\ddots   \\
 &&\bV_{\Nc}^{H}
\end{array}\right]}_{\triangleq{\tilde{\bV}}}\bQ^{-1}\Fbb}_{\triangleq{\bV^{H}}},
\end{align}
The GEC-SR algorithm can be performed by matrix-vector multiplications with $\bU$, $\bS$ and $\bV^{H}$. Therefore, the diagonal elements of $\bQ_{2\bx}$ and $\bQ_{2\bz}$ are expressed as
\begin{equation}\label{dQ2x}
  q_{2\bx}=\frac{1}{N}\sum_{i=1}^{N}\left({\frac{1}{v_{2x}}+\frac{s_{i}^{2}}{v_{2z}}}\right)^{-1},
\end{equation}

\begin{equation}\label{dQ2z}
  q_{2\bz}=\frac{1}{M}\sum_{i=1}^{N}s_{i}^{2}\left({\frac{1}{v_{2x}}+\frac{s_{i}^{2}}{v_{2z}}}\right)^{-1},
\end{equation}
because we only focus on diagonal elements to compute extrinsic information.
Furthermore, the corresponding posterior mean estimate of $\bx$ in lines $6$ and $14$ can be derived by
\begin{equation}\label{psotxSVD}
  \hat{\bx}_{2}=\bV\bE_{t}\left(\frac{\bV^{H}\br_{2\bx}}{v_{2x}}+\frac{\bS^{H}\bU^{H}\br_{2\bz}}{v_{2z}}\right),
\end{equation}
where $\bE_{t}$ is an $N\times N$ diagonal matrix such that
\begin{equation}\label{Singular_matrix}
[\bE_{t}]_{nn}=\left(\frac{1}{v_{2x}}+\frac{s_{i}^{2}}{v_{2z}}\right)^{-1}.
\end{equation}

The computational complexity order for the matrix inversion in lines $5$ and $13$ is $\mathcal{O}(MN^{2})$, which is highly complex for large-dimensional matrix. Conversely, the computational complexity of the OFDM-based method is dominated by $(\ref{psotxSVD})$. Notably, in $(\ref{psotxSVD})$, $\bV^{H}\br_{2\bx}$ is given by
\begin{equation}\label{v1r2x}
\bV^{H}\br_{2\bx}=\tilde{\bV}\bQ^{-1}\Fbb\br_{2\bx},
\end{equation}
where $\Fbb\br_{2\bx}$ can be performed using simple FFT operation. Meanwhile, $\bQ$ is a permutation matrix to implement corresponding columns exchange  that does not cost any computational resource. $\tilde{\bV}$ is a block diagonal matrix that is composed of $\bV_{i}^{H}$; therefore, corresponding matrix-vector multiplications can be implemented through these block matrices $\bV_{i}^{H}$. Furthermore, $\bS^{H}\bU^{H}\br_{2\bz}$ can be realized using the same idea to reduce complexity. The computational complexity of the OFDM-based method is $\mathcal{O}(2N\log_{2}N+M\log_{2}\Nc+2\Nct N+2N+\Ncr M)$, which is much smaller than $\mathcal{O}(MN^{2})$, especially for a large number of subcarriers. Thus, the computational complexity is significantly decreased. This low-complexity realization is performed only by matrix-vector multiplications and FFT operation which are highly suitable for hardware implementation. Furthermore, the complexity of the GAMP algorithm, which is the representative signal reconstruction algorithm, is $\mathcal{O}(MNt_\mathrm{max})$ \cite{Zhang2016TWC}. $t_{\mathrm{max}}$ is the iteration number, and the GAMP algorithm converges within  $10$ iterations in most cases. Therefore, the computational complexity of this OFDM-based realization is much smaller than that of the GAMP algorithm. We will compare the performance of the GAMP and the the GEC-SR algorithms in the simulation.


\subsection{Mixed-ADC Detection}\label{Mixed-ADC}
Our GEC-SR algorithm is very flexible to various architectures. For a mixed-ADC architecture in which low-resolution and high-resolution ADCs are simultaneously adopted at the receiver, only module A need to be revised. Therefore, the optimal detector is easily extended to a mixed-ADC architecture.
 By considering high-resolution RF chains as AWGN channels, we obtain the likelihood function
\begin{equation}\label{unquantized}
  \sfP_{\mathrm{out}}(\tilde{y}|z)=\frac{1}{\pi\sigma_{N}^{2}}e^{-|\tilde{y}-z|^{2}/\sigma_{N}^{2}}.
\end{equation}
\vspace{0cm}
To distinguish the various ADCs, we use $\Omega_{\kappa}$ to indicate the collection of RF chains equipped with ADCs of $\kappa$ bits. Moreover, the cardinality of $\Omega_\kappa$ is $N_{r,\kappa}$, and thus we have $\sum_{\kappa}N_{r,\kappa}=N_{r}$. For high-resolution RF chains, the explicit expressions of the posterior mean and variance are given by
\begin{align}
    \hat{z}_{1,\infty}
   &= r_{1z} + \frac{v_{1z}}{v_{1z}+\sigma_{N}^{2}}(\tilde{y}-r_{1z}),
   \label{finiteADCmean} \\
    v_{1z,\infty}^{\mathrm{post}} & =v_{1z}-\frac{v_{1z}^{2}}{v_{1z}+\sigma_{N}^{2}} ,
     \label{finiteADCvar}
\end{align}
Therefore, for mixed-ADC architecture, line $1$ computes the posterior mean by using respective analytical expression in (\ref{eq:hatZ_RealGaussian}) and (\ref{finiteADCmean}), and line $2$ should consider the contribution of low-resolution and high-resolution ADCs in (\ref{eq:mseZ_RealGaussian}) and (\ref{finiteADCvar}), as given by
\begin{equation}\label{mixedvar}
  \mathbf{v}_{1\mathbf{z}}^{\mathrm{post}}=[v_{1z,\kappa}^{\mathrm{post}}\mathbf{1}^T_{N_{r,\kappa}},v_{1z,\infty}^{\mathrm{post}}\mathbf{1}^T_{N_{r,^\infty}}]^T
\end{equation}
where $v_{1z,\kappa}^{\mathrm{post}}$ is the posterior variance that is computed by (\ref{eq:mseZ_RealGaussian}).

\section{Performance Analysis}\label{performance analysis}
In this section, we show that the asymptotic performance of the proposed algorithm can be characterized by the recursion of a set of state evolution (SE) equations. We focus on a general mixed-ADC architecture, and pure low-resolution architecture is a special case of mixed-ADC system. Our derivation is performed in the large-system regime where $M$ and $N$ tend to infinity, whereas the ratio
\begin{equation}\label{largesystem}
M/N\rightarrow\beta,~~N_{r,\kappa}/N_{r}\rightarrow\beta_{\kappa},~~ \forall \kappa
\end{equation}
remains fixed.
We first show the SE equations in Proposition \ref{proposition1} and subsequently give detailed explanation for each equation linked with Algorithm 1.


Proposition \ref{proposition1} involves several parameters that can be illustrated systematically by a scalar AWGN channel
\begin{equation}\label{scalar_AWGN}
  r=x+w,
\end{equation}
where $w\sim\cN_{\bbC}(x;0,v_{1x})$. The posterior mean estimator for $x$ reads
\begin{equation}\label{psterior_x}
  \hat{x}=\int {x} \sfP(x|{r})\mathrm{d}x,
\end{equation}
and thus the MSE of the estimators are given by
\begin{equation}\label{MSE_x}
 \mathsf{MSE}_{\rmx}=\sfE\{|x-\hat{x}|^{2}\},
\end{equation}
where the expectations are taken over $\sfP(r,x)$, the specific expression of $\mathsf{MSE}_{\rm x}$ is dependent on the distribution of signal $x$.
\begin{proposition}\label{proposition1}
In the large-system limit, the asymptotic behavior of the algorithm can be described by a set of equations.
\begin{subequations} \label{state_evolution}
   \begin{align}
     &\alpha_{\kappa}^{t+1}=\frac{1}{2}\sum_{\tilde{y} \in \kappa}\int Du\frac{\bigg(\Psi'\left(\tilde{y};\sqrt{\frac{v_{z}-v_{1z}^{t}}{2}}u, \frac{\sigma_{N}^2 +v_{1z}^{t}}{2} \right)\bigg)^{2}}{\Psi\left(\tilde{y};\sqrt{ \frac{v_{z}-v_{1z}^{t}}{2}}u, \frac{\sigma_{N}^2 +v_{1z}^{t}}{2} \right)}\label{SEeqa} \\
     &v_{1z,\infty}^{\mathrm{post},t+1}=\frac{v_{1z}^{t}\sigma_{N}^{2}}{v_{1z}^{t}+\sigma_{N}^{2}}\label{SEeqb}\\
     &v_{1z,\kappa}^{\mathrm{post},t+1}=v_{1z}^{t}-\alpha_{\kappa}^{t+1}(v_{1z}^{t})^{2}\label{SEeqc}\\
     &v_{1z}^{\mathrm{post},t+1}=\beta_{\kappa}v_{1z,\kappa}^{\mathrm{post},t+1}+(1-\beta_{\kappa})v_{1z,\infty}^{\mathrm{post},t+1}\label{SEeqd}\\
     &\gamma_{2z}^{t+1}=\frac{1}{v_{1z}^{\mathrm{post},t+1}}-\gamma_{1z}^{t}\label{SEeqe}  \\
     &q_{x}^{t+1}=\sfE\Bigg\{\frac{1}{\lambda_{i}\gamma_{2z}^{t+1}+\gamma_{2x}^{t}}\Bigg\}\label{SEeqf} \\
     &\gamma_{1x}^{t+1}=\frac{1}{q_{x}^{t+1}}-\gamma_{2x}^{t}\label{SEeqg} \\
     &\gamma_{2x}^{t+1}=\frac{1}{\mathsf{MSE}_{\rmx}(\gamma_{1x}^{t+1})}-\gamma_{1x}^{t+1}\label{SEeqh} \\
     &q_{z}^{t+1}=\sfE\Bigg\{\frac{\lambda_{i}}{\lambda_{i}\gamma_{2z}^{t+1}+\gamma_{2x}^{t+1}}\Bigg\}\label{SEeqi} \\
     &\gamma_{1z}^{t+1}=\frac{1}{q_{z}^{t+1}}-\gamma_{2z}^{t+1}\label{SEeqj},
   \end{align}
  \end{subequations}
where $t=0,1,2,\ldots$ denotes the iteration index, $v_{z}=\sfE\{\lambda\}\mathsf{E}\{|x_{n}|^{2}\}$, the initialization $v^{0}_{1z}=\frac{N}{M}\sfE\{\lambda\}$, $\gamma_{2x}^{0}=1$, and
\begin{align}
		&\Psi\left(\tilde{y};z,u^2\right)\triangleq\Phi\left(\frac{z-r^{\mathrm{low}}}{u}\right)-\Phi\left(\frac{z-r^{\mathrm{up}}}{u}\right),\displaybreak[0]\notag\\
		&\Psi'\left(\tilde{y};z,u^2\right)\triangleq\frac{\partial\Psi\left(\tilde{y};z,u^2\right)}{\partial z}=\frac{\phi\left(\frac{z-r^{\mathrm{low}}}{u}\right)-\phi\left(\frac{z-r_{\mathrm{up}}}{u}\right)}{ u }\notag.
\end{align}
\hfill\ensuremath{\blacksquare}
\end{proposition}
For ease of expressions, we define several auxiliary parameters as follows:
\begin{equation}\label{new_parameters}
\gamma_{1z}=\frac{1}{v_{1z}}, ~~\gamma_{2z}=\frac{1}{v_{2z}}, ~~\gamma_{1x}=\frac{1}{v_{1x}},~~\mathrm{and} ~~\gamma_{2x}=\frac{1}{v_{2x}}.
\end{equation}
In addition, Proposition \ref{proposition1} involves expectation operator w.r.t. the distribution of eigenvalue, where $\lambda_{i}$ is the $i$-th eigenvalue of $\bA^{H}\bA$, the expectation w.r.t. $\lambda$ is defined by
\begin{equation}\label{expectation_lamda}
\sfE\{f(\lambda)\}=\frac{1}{N}\sum\limits_{i=1}^{N}f(\lambda_{i}).
\end{equation}

Although a proof for Proposition \ref{proposition1} is provided in Appendix A, we give an intuitive explanation for Proposition \ref{proposition1}, which starts from the GEC-SR algorithm itself. SE equations can be deduced from Algorithm 1 in each step. In module A, we determine the asymptotic behaviors of $\hat{z}_{1}$ by its associated variance $v_{1z}^{\mathrm{post}}$, which is computed by (\ref{SEeqa})-(\ref{SEeqd}) and are identical to line $2$. We exploit a mixed-ADC architecture in which the contribution of high-resolution and low-resolution ADCs should be considered simultaneously. Therefore, (\ref{SEeqd}) can be interpreted as the weighted average of $v_{1z,\kappa}^{\mathrm{post}}$ and $v_{1,z,\infty}^{\mathrm{post}}$, $\beta_{\kappa}$ and $1-\beta_{\kappa}$ represent the ratios of low-resolution and high-resolution ADCs, respectively.
In module C, lines $5$, $13$, and $15$ aim to compute the posterior covariance matrix $\bQ_{2\bx}$ and $\bQ_{2\bz}$. Only the diagonal elements of the $\bQ_{2\bx}$ and $\bQ_{2\bz}$ are needed, which is obtained by (\ref{SEeqf}) and (\ref{SEeqi}). In module B, when computing $\mathsf{MSE}_{\rmx}$, (\ref{SEeqh}) involves the MMSE estimate of $x$ under an AWGN corrupted observation, which is identical to line $10$ and (\ref{scalar_AWGN}).

Before proceeding, we explain several characteristics of Proposition \ref{proposition1}. We also provide two examples to understand SE equations explicitly.

\emph{1) Theoretical Tractability}: From (\ref{AWGN_estimate}) and (\ref{scalar_AWGN}), we observe that Algorithm 1 and Proposition \ref{proposition1} can be characterized by equivalent AWGN model with equivalent SNR $\frac{1}{v_{1x}}=\gamma_{1x}$, and the scalar model is independent between each symbol $x_{i}$. This characteristic indicates that Bayesian optimal data detector is decoupled into $N$ uncoupled scalar equivalent model (\ref{scalar_AWGN}). Using the scalar equivalent channel, we can easily predict several fundamental performance metrics, such as MSE, SER, and mutual information. We provide two special cases for Proposition \ref{proposition1} as follows:

{\noindent {\bf Example~1} (Constellation-like Inputs).} Based on Proposition \ref{proposition1} , the asymptotic MSEs can be determined using the MSEs of the scalar AWGN channels (\ref{scalar_AWGN}). Thus, if the data symbol is drawn from a quadrature phase-shift keying (QPSK) constellation, then we will derive
\begin{equation}\label{mse_x_explicit expression}
  \mathsf{MSE}_{\rm x}=1-\int \rmD z\tanh(\gamma_{1x}+\sqrt{\gamma_{1x}}z)
\end{equation}
The SER w.r.t. $\bx$ can also be evaluated through the scalar AWGN channel (\ref{scalar_AWGN}), which is given by \cite{CKWen2016TSP}.
\begin{equation}\label{SER_QPSK}
  \mathrm{SER}=2Q(\sqrt{\gamma_{1x}})-[Q(\sqrt{\gamma_{1x}})]^{2}
\end{equation}
where $Q(x)= \int_{x}^{\infty}Dz$ is the Q-function. All these performances can be determined on the basis of knowledge of the scalar AWGN channel with SNR $\gamma_{1x}$ (\ref{scalar_AWGN}), which is known as the decoupling principle. Thus, if the data symbol is drawn from other square QAM constellations, then the corresponding SER can be easily obtained using the closed-form SER expression in \cite{Proakis2007}.

{\noindent {\bf Example~2} (Unquantized Channel).} In MIMO-OFDM system with infinite-resolution ADCs, $\Nc$ subcarriers are mutually orthogonal.
In each subcarrier, assuming equal power allocation, a $\Ncr\times\Nct$ MIMO channel can be decomposed into $\Nct$ parallel data streams. Therefore, we can obtain $\Nct\Nc$ parallel subchannels completely with identical SNR, which is $\sfE\{\lambda\}/\sigma_{N}^{2}$. Owing to severe distortion introduced by low-resolution ADCs, the orthogonality among subcarriers is destroyed. However, Proposition \ref{proposition1} and (\ref{AWGN_estimate}) demonstrate that quantized MIMO-OFDM can still be decoupled into $\Nct\Nc$ equivalent AWGN subchannels:
\begin{equation}\label{decoupling_principle}
  r_{i}=\sqrt{\gamma_{1x}}x_{i}+w_{i}
\end{equation}
for $i=1,\ldots,N$, where $w_{i}\sim\cN_{\bbC}(0,1)$, and $\gamma_{1x}$ is the equivalent SNR for quantized MIMO-OFDM systems.

Next, we show that as infinite-resolution ADCs are employed, $\gamma_{1x}$ in \eqref{decoupling_principle} for quantized MIMO-OFDM can reduce to $\sfE\{\lambda\}/\sigma_{N}^{2}$ for unquantized MIMO-OFDM.
To this end, we let $r^{\mathrm{low}} = r$ and $r^{\mathrm{up}} = r^{\mathrm{low}} + \rmd r$. As $\rmd r \to 0$, we obtain
$\Phi\left(\frac{z-r^{\mathrm{low}}}{u}\right)-\Phi\left(\frac{z-r^{up}}{u}\right) \to \frac{\rm d}{\rmd r}
\Phi\left(\frac{z-r}{u}\right)$ and $\phi\left(\frac{z-r^{\mathrm{low}}}{u}\right)-\phi\left(\frac{z-r^{\mathrm{up}}}{u}\right) \to \frac{\rm d}{\rmd r}
\phi\left(\frac{z-r}{u}\right)$. By substituting these relationships into (\ref{SEeqa}) and applying the facts that $\frac{\rm d}{\rmd r}
\Phi\left(\frac{z-r}{u}\right) = \frac{1}{u} \phi\left(\frac{z-r}{u}\right)$ and $\frac{\rm d}{\rmd r} \phi\left(\frac{z-r}{u}\right) =
\left(\frac{z-r}{u^2}\right) \phi\left(\frac{z-r}{u}\right)$, we can obtain

\begin{equation}\label{alpaht}
     \alpha^{t}=\frac{1}{\sigma_{N}^{2}+v_{1z}^{t}},
\end{equation}
and substituting (\ref{alpaht}) into Proposition \ref{proposition1} , we have
\begin{equation}\label{v1ztinf}
   v_{1z,\kappa\rightarrow\infty}^{\mathrm{post},t}=v_{1z}^{t}-\alpha^{t}(v_{1z}^{t})^{2}=\frac{v_{1z}^{t}\sigma_{N}^{2}}{v_{1z}^{t}+\sigma_{N}^{2}},
\end{equation}
which is consistent with (\ref{finiteADCvar}). Finally, the equivalent SNR is $\gamma_{1x}=\sfE\{\lambda\}/\sigma_{N}^{2}$, which exactly matches unquantized MIMO-OFDM systems.
This agreement also indicates accuracy of SE equations.

\emph{2) Optimality}: 
From the statistical mechanics perspective \cite{Shinzato2009JPA,Krzakala2012JSM}, the performance metrics of the Bayesian MMSE estimator, such as MSE, is equivalent to finding the saddle points of the free energy defined by
\begin{equation}\label{free_energy}
  \cF=-\frac{1}{M}\rm\sfE\{\log p(\tilde{\by})\}.
\end{equation}
The calculation of $\cF$ is very difficult. Fortunately, the replica method from statistical physics provides a highly sophisticated procedure to address this calculation. The calculation of $\cF$ and its saddle point are given in Appendix B. Interestingly, the iterative procedure of the GEC-SR algorithm is equivalent to finding the saddle points of the free energy. This phenomenon indicates that the proposed detector can achieve optimal detection performance in MMSE sense. Therefore, our proposed detector can be regarded as a lower bound in SER and MSE performance for other detectors in quantized MIMO-OFDM systems.

\emph{3) Generality and Computation Simplicity}: Proposition 1 incorporates the effect of mixed-ADC architecture, and arbitrary quantization processes, such as nonuniform quantization and different quantization levels in (\ref{SEeqa})-(\ref{SEeqd}). Furthermore, different prior distribution $\sfP(\bx)$ is also involved in these SE equations. The analytical result is computationally tractable because the corresponding parameters can be obtained iteratively. Therefore, instead of time-consuming Monte Carlo simulations to evaluate mixed-ADC performance, we can predict theoretical behavior by SE equations.
\section{Simulation Results and Discussions}\label{Simulation Results and Discussions}
\subsection{Accuracy of the Analytical Results}\label{Analytical Results}
We will present numerical simulation results for the proposed Bayesian optimal data detector in mmWave MIMO-OFDM systems with low-resolution ADCs. In the simulations, we assume the number of channel taps is $L=4$. The channel impulse response $h_{n_{r}n_{t},i}$ for $i=1,\ldots,L$ is assumed to be i.i.d. with PDF $\cN_{\bbC}(h_{n_{r}n_{t},i};0,1/L)$. Each entry of the transmitted symbols $\bx$ is drawn from the equiprobable QPSK constellation without specific indication. We set $E[|x_{j}|^{2}]=1$ for $j=1,\ldots,N$, and the channel matrix $\bA$ is normalized by divided $\sqrt{\Nct}$, thus the average SNR can be given by $1/\sigma_N^{2}$. The SERs and MSEs, which are averaged over all data streams, are obtained through the Monte Carlo simulations of $10,000$ independent channel realizations. We focus on small systems with $\Ncr=\Nct=N_{s}=2$ due to the small number of RF chains in mmWave systems.
\begin{figure}[!t]\centering	
\subfloat[$\Nc=1024$, QPSK symbol, SNR=10dB]{\includegraphics[width=3in]{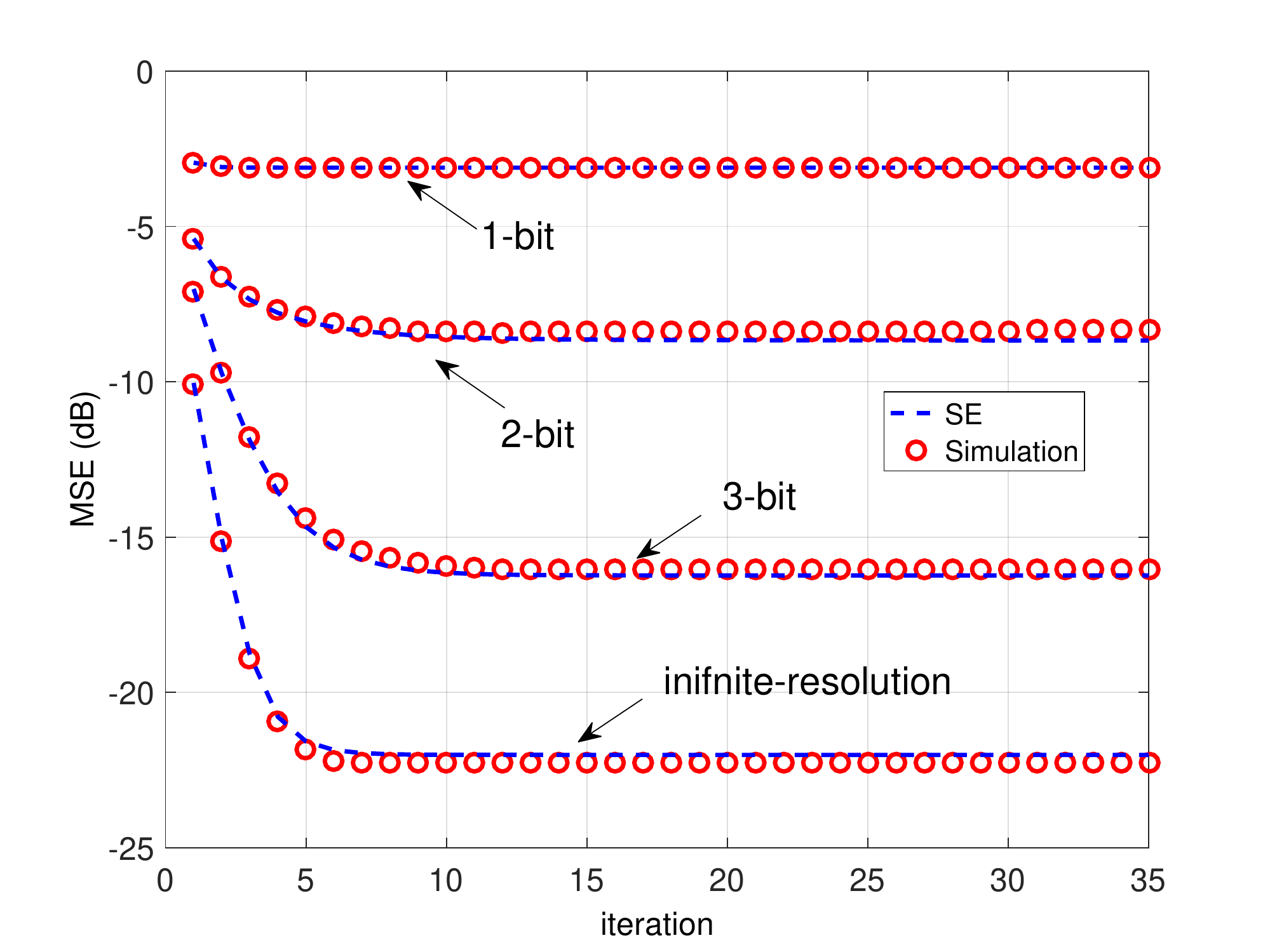}\label{fig3a}}\\
	\subfloat[$\Nc=64$, QPSK symbol]{\includegraphics[width=3in]{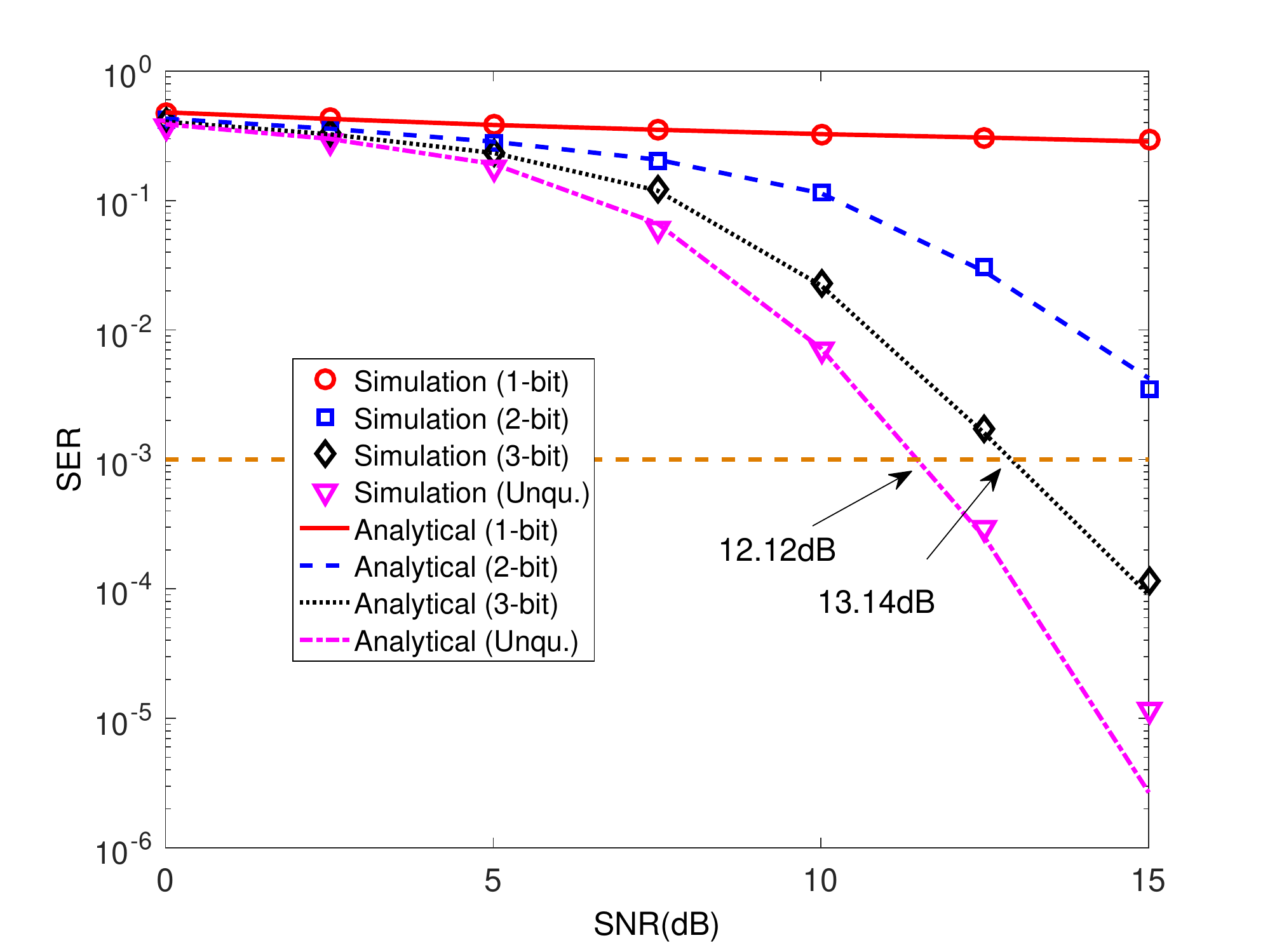}\label{fig3b}}
	\caption{.~~MSEs and SERs performance of the proposed detector under different quantization levels.
}\label{fig3}
\end{figure}

Fig. \ref{fig3} shows the MSEs and SERs of the proposed detector, under the quantization precision of $1$-$3$ bits and high-resolution ADCs. The simulated results are obtained by the Monte Carlo simulations of the GEC-SR algorithm, whereas the analytical results are evaluated using SE equations.
Fig. \ref{fig3a} demonstrates that the proposed detector evidently converges within five iterations, and such a convergence is rapid. We also observe that the GEC-SR algorithm can generally describe the performance of theoretical Bayes-optimal estimator in each iteration.
Fig. \ref{fig3b} illustrates that the SERs of the proposed detector match well with theoretical results when time index $t\rightarrow t_\mathrm{max}$. Interestingly, the SE predictions which is derived in the large system limit where $N\rightarrow\infty$, are precise even for a small number of subcarriers. Therefore, instead of performing time-consuming Monte Carlo simulations to obtain the corresponding performance metrics, we can predict the theoretical behavior by SE equations in a very short time. In addition, we observe that with the increase of quantization precision, the proposed detector achieves significant performance improvement, and the performance degradation due to low-precision quantization is small when quantization precision is $3$-bit. For example, if we target the SNR that attained by the unquantized system at $\mathrm{SER}=10^{-3}$, then the $3$-bit
quantization incurs a loss of $13.14-12.12=1.02$ dB, which remains acceptable in the quantized mmWave MIMO-OFDM system design. This result illustrates the feasibility of using low-resolution ADCs at mmWave MIMO-OFDM systems.

\begin{figure}
  \centering
  \includegraphics[width=3in]{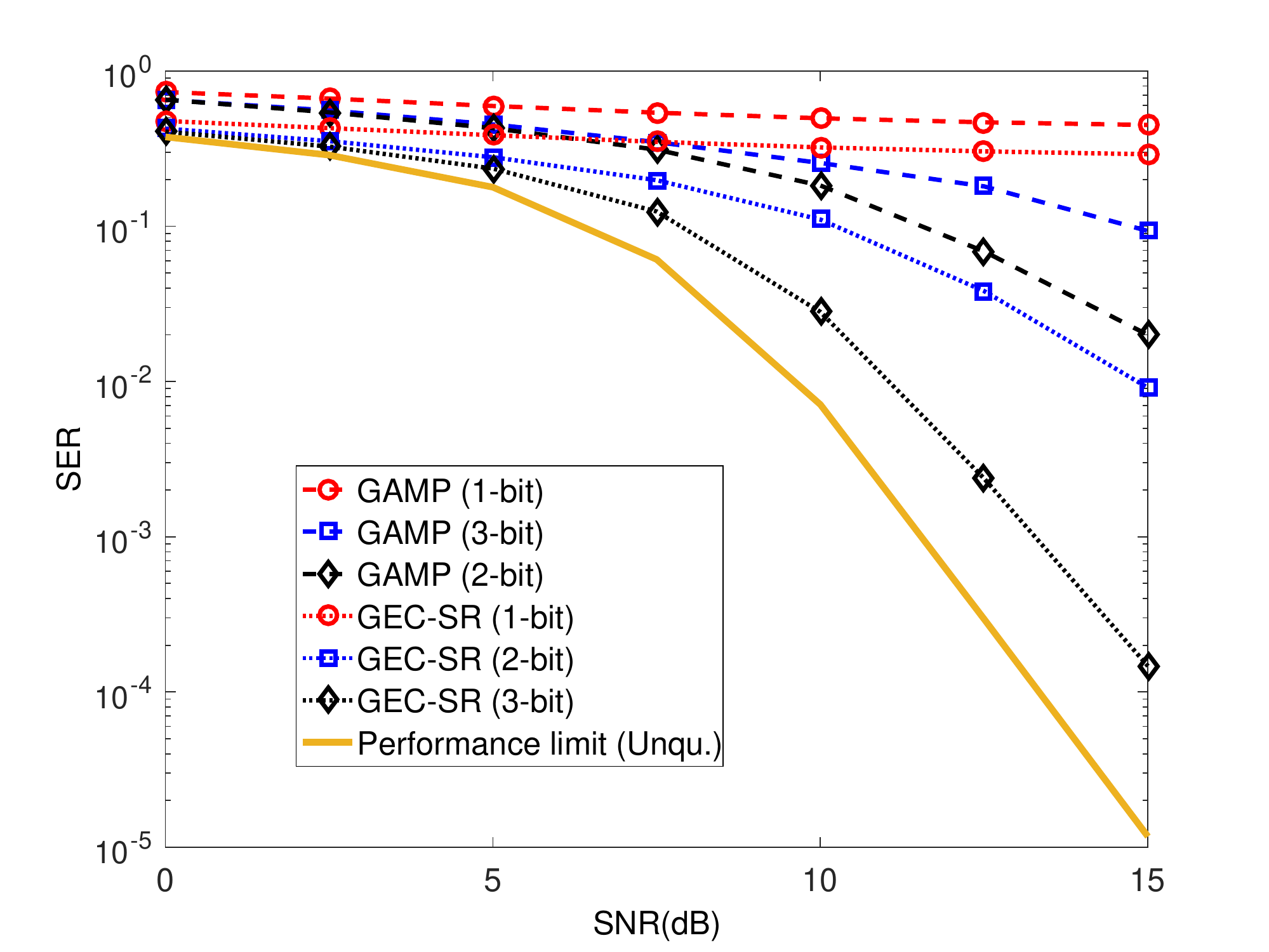}
  \caption{.~~SERs performance comparisons of the proposed detector and GAMP-based detector with $\Ncr=\Nct=N_{s}=2$ under different quantization levels. }\label{fig4}
\end{figure}


As is shown in \cite{Zhang2016TWC} that LMMSE detector perform worse than GAMP-based detector under quantized systems because the former ignores the de-quantization operation. Therefore, instead of ordinary LMMSE detector, we choose GAMP-based detector for comparison. Fig. \ref{fig4} compares the SERs of the proposed GEC-based detector and that of GAMP-based detector. Notably, the proposed detector significantly outperforms GAMP-based detector in terms of SER performance. We observe the poor performance obtained by the GAMP detector, because GAMP algorithm is fragile in terms of the choice of channel matrix, and thus perform poorly in quantized MIMO-OFDM channel.

\subsection{Mixed-ADC Architecture}\label{mixed-ADC architecture}
We have discussed the optimal data detector and theoretical analysis under pure low-resolution ADCs, and extended to mixed-ADC architecture. Based on this optimal data detector, we investigate the performance of the mixed-ADC architecture. Our discussions focus on two cases. One is replacing a few low-resolution ADCs with high-resolution ADCs, and the other is adding a few low-resolution RF chains in original high-resolution ADC architecture. These two cases represent the effect of high-resolution ADCs and low-resolution RF chains.
\begin{figure}[!t]\centering	
\subfloat[Mixed $1$-bit architecture]{\includegraphics[width=3in]{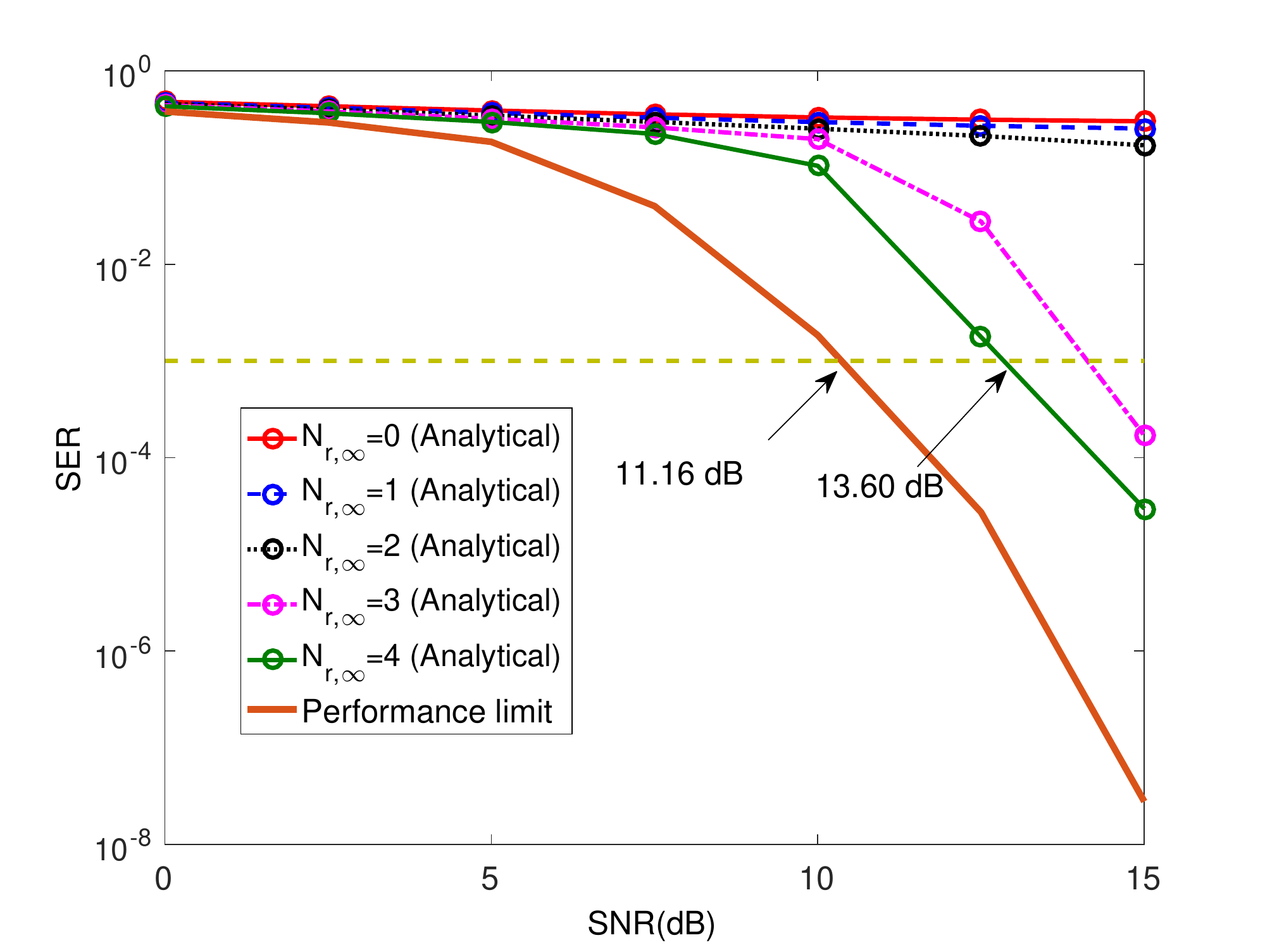}\label{fig5a}}\\
	\subfloat[Mixed $2$-bit architecture]{\includegraphics[width=3in]{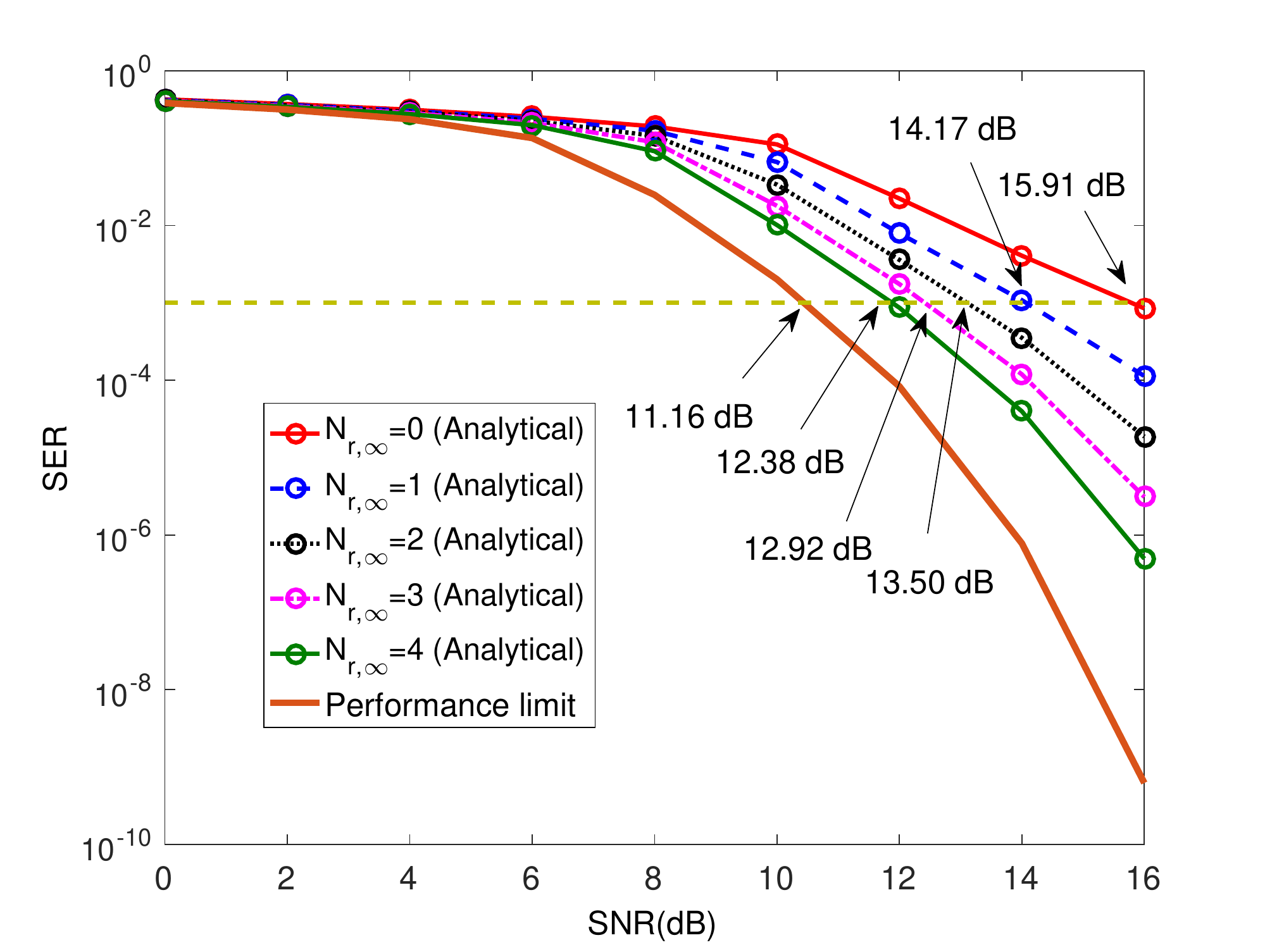}\label{fig5b}}
	\caption{.~~SERs versus SNR under the mixed-ADC architecture for different number of high-resolution ADCs with QPSK inputs.}\label{fig5}
\end{figure}


\emph{1) High-resolution ADCs:} Fig. \ref{fig5} illustrates the performance of the mixed-ADC architecture.  $N_{r,\infty}$ denotes the number of high-resolution ADCs, and $N_{r,\kappa}$ is the number of $\kappa$-bit low-resolution ADCs where $N_{r,\infty}+N_{r,\kappa}=N_{r}$. In this simulation, we set $\Ncr=\Nct=N_{s}=8$, and the number of subcarrier $\Nc=128$. From (\ref{Analytical Results}), we observe that when the number of RF chains in receiver is equal to that in transmitter, pure low-resolution ADCs architecture will arouse severe distortion, especially for $1$-bit ADC. This outcome differs from typical massive MIMO systems (e.g., $\Ncr/\Nct=10$), where performance loss caused by low-resolution ADCs can be compensated by increasing the
number of receiving antennas. In this case, employing high-resolution ADCs that turn into mixed-ADC architecture is an efficient means to decrease distortion. Our discussion turns to the following question: \emph{how many high-resolution RF chains do we need to render quantization distortion acceptable?} As shown in Fig. \ref{fig5a}, for mixed $1$-bit architecture, with less than two high-resolution ADCs installed, the performance improvement is poor. When three RF chains are adopted in the receiver, the performance improves drastically, and the error floor is eliminated. This result is because serious non-linear distortion is involved under $1$-bit quantization, and the amplitude information of received signal is completely lost. In this case, we need more high-resolution ADCs to assist data detection. Compared with the performance limit, installing four high-resolution ADCs only exhibits $13.60-11.16=2.44$ dB loss.


   Fig. \ref{fig5b} investigates the potential of high-resolution ADCs under the mixed $2$-bit architecture. We observe that SER performance generally improve $15.91-14.17=1.74$ dB by equipping one high-resolution ADC. Furthermore, two high-resolution ADCs can bring $15.91-13.50=2.41$ dB, and for three high-resolution ADCs, the gain is $15.91-12.92=2.99$ dB. 
    Compared with the performance limit, installing four high-resolution ADCs only displays $12.38-11.16=1.22$ dB loss.
 \begin{figure}[t]
  \centering
  \includegraphics[width=3in]{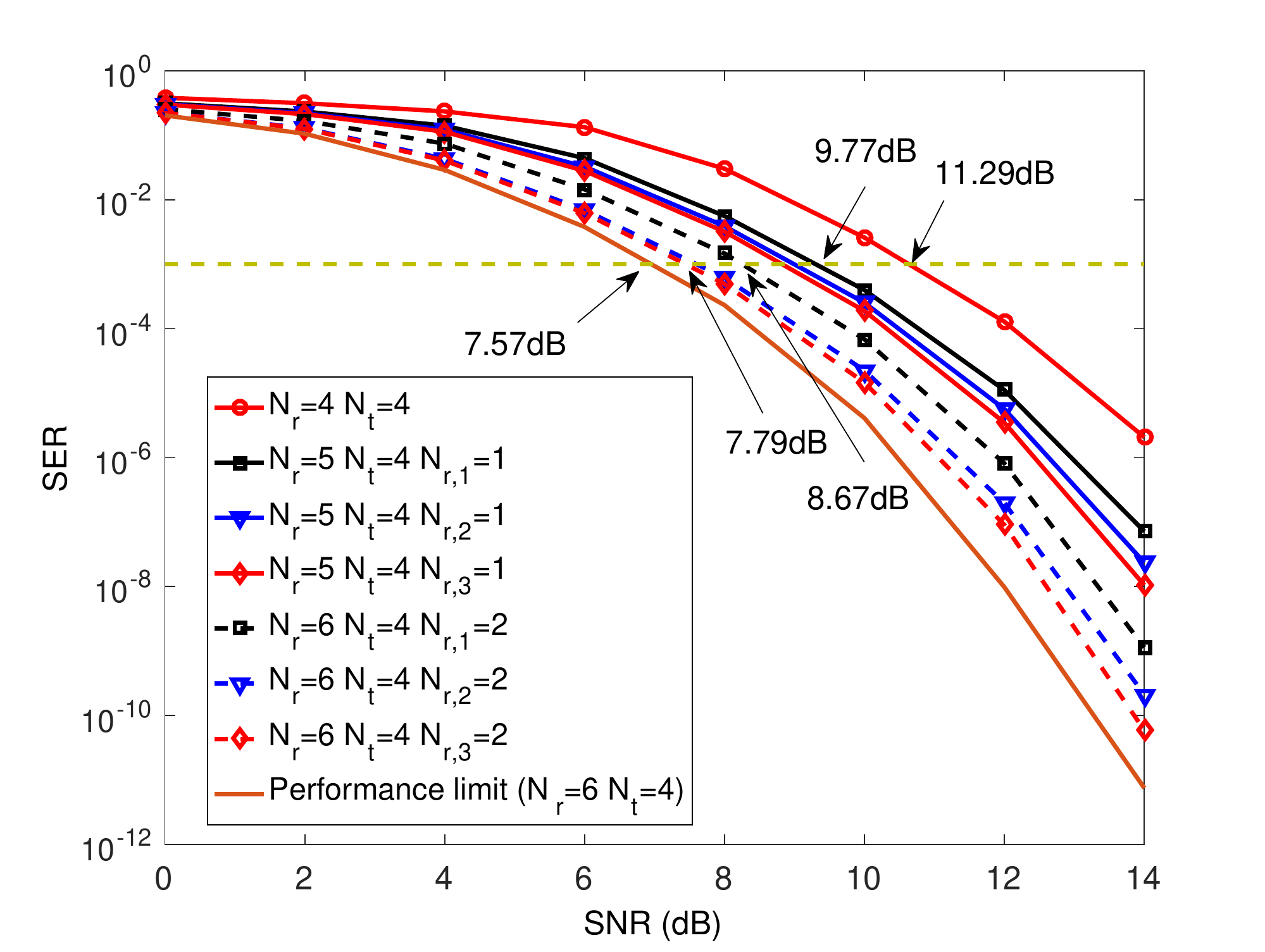}
  \caption{.~~SERs versus SNR for mixed-ADC architecture by adding different numbers of low-resolution RF chains with QPSK inputs.}\label{fig6}
\end{figure}

\emph{2) Low-resolution RF chains:} Fig. \ref{fig6} shows the SER performance of mixed-ADC architecture by adding different number of low-resolution RF chains. We set $N_{t,\infty}=N_{r,\infty}=N_{s}=4$, $N_{r,\kappa}$ denotes the number of low-resolution RF chains, and the number of subcarrier $\Nc=128$. Our discussions focus on the following question: \emph{how much gain can we obtain by adding a few low-resolution RF chains over original high-resolution architecture?} Fig. \ref{fig6} demonstrates that when we target SER=$10^{-3}$, by adding one $1$-bit RF chain, the SER performances improve by $11.29-9.77=1.52$ dB. By contrast, two $1$-bit RF chains can bring $11.29-8.67=2.62$ dB gain.
 We observe that when two $3$-bit RF chains are added, only a gap of $7.79-7.57=0.22$ dB is noted compared with performance limit. Interestingly, the performance improvement of increasing the number of $1$-bit RF chains is more effective than that of increasing the bit resolution of ADCs. Therefore, adding a few low-resolution RF chains is useful to improve the system performance, and the low-resolution RF chain is hardware-friendly, especially for $1$-bit RF chain.

\section{Conclusions}\label{conclusions}
We developed a framework for studying Bayesian optimal data detector in hybrid broadband mmWave systems with low-resolution ADCs, and extended it to mixed-ADC architecture. Importantly, we proposed an efficient algorithm to achieve Bayes-optimal data detection by applying GEC-SR technique, which is computationally tractable. The asymptotic behaviors of the performance metric (MSEs and SERs) were derived on the basis of SE equations which were identical to those obtained by the replica method. This consistency indicated the optimality of the data detector in MMSE sense. Monte Carlo simulations demonstrated the accuracy of our analytical results. Furthermore, the accurate analytical expressions enabled us to quickly and efficiently evaluated the performance of mmWave quantized MIMO-OFDM systems. We also provided useful observations as insights into the system design of a mixed-ADC architecture. Our results show that adding a few low-resolution RF chains in original high-resolution system can obtain significant gain such as two $1$-bit RF chains can bring $2.62$ dB gain. Adopting high-resolution ADCs was also useful to reduce performance gap and eliminate error floor.
 \section*{Appendix A: proof of proposition 1}\label{AP1}

To prove Proposition \ref{proposition1}, first, we should obtain the large-system behavior of $v_{1z}^{\mathrm{post}}$ in module A, which can be derived by using a strategy similar to those presented in \cite{wang2017Gturbo} and \cite{Liu2016ISIT}. The final expression of the $v_{1z,\kappa}^{\mathrm{post}}$ for the quantized channel is given by \cite{wang2017Gturbo}
\begin{align}\label{v1zpost2}
\mathsf{E}{\left[v^{\mathrm{post}}_{1z,\kappa}\right]}&=v_{1z}-\frac{(v_{1z})^2}{2}\times    \nonumber \\
&\sum\limits_{\tilde{y}\in\cR_{{\kappa}}}\int_{-\infty}^{\infty}\frac{{\left[\Psi'{\left(\tilde{y};\sqrt{\frac{v_z-v_{1z}}{2}}z,\frac{\sigma_{N}^2+v_{1z}}{2}\right)}\right]}^2}{\Psi{\left(\tilde{y};\sqrt{\frac{v_z-v_{1z}}{2}}z,\frac{\sigma_{N}^2+v_{1z}}{2}\right)}}\mathrm{D}z.
\end{align}
Considering the contributions of various low-resolution and high-resolution ADCs, the asymptotic behavior of $v_{1z}^{\mathrm{post}}$ is obtained as the weighted average of $v_{1z,\kappa}^{\mathrm{post}}$ and $v_{1,z,\infty}^{\mathrm{post}}$, which is given by
\begin{equation}\label{VAR_mixed}
v_{1z}^{\mathrm{post}}=\beta_{\kappa}v_{1z,k}^{\mathrm{post}}+(1-\beta_{\kappa})v_{1z,\infty}^{\mathrm{post}}.
\end{equation}

After computing the extrinsic information, we can obtain \emph{cavity} variance $\bv_{2\bz}$.
In a large system limit, module A passes the extrinsic information that can be regarded as Gaussian random variable denoted by $\bz_{2}\sim\cN_{\bbC}(\bz_{2};\br_{1\bz},\mathrm{Diag}(\bv_{2\bz}))$, and module B passes the extrinsic information that can be regarded as Gaussian random variable denoted by $\bx_{2}\sim\cN_{\bbC}(\bx_{2};\br_{1\bx},\mathrm{Diag}(\bv_{2\bx}))$. The sum-product belief is $b_{sp}\propto \cN_{\bbC}(\bx_{2};\br_{2\bx},\mathrm{Diag}(\bv_{2\bx}))\cN_{\bbC}(\bA\bx_{2};\br_{2\bz},\mathrm{Diag}(\bv_{2\bz}))$ under the restriction $\bz_{2}=\bA\bx_{2}$. Using standard Gaussian integral identities, we can show that this belief is also Gaussian with $\bx_{2}\sim\cN_{\bbC}(\bx_{2}; \hat{\bx}_{2},\bQ_{2\bx})$, and its mean and variance are given by
\begin{align}
  \bQ_{2\bx} &= (\mathrm{Diag}(\mathbf{1}\oslash\bv_{2\bx})+\bA^{H}\mathrm{Diag}(\mathbf{1}\oslash\bv_{2\bz})\bA)^{-1}\label{beliefmeanx}\\
  \hat{\bx}_{2} & = \bQ_{2\bx}(\br_{2\bx}\oslash\bv_{2\bx}+\bA^{H}\br_{2\bz}\oslash\bv_{2\bz})\label{beliefvarx}.
\end{align}
which can be interpreted as LMMSE estimate of $\bx$ under the following assumption:
\begin{equation}\label{LMMSE_estimate1}
  \br_{2\bz}=\bz_{2}+\bw_{2\bz},
\end{equation}
where $\bw_{2\bz} \sim \cN_{\bbC}(\mathbf{0},\mathrm{Diag}(\bv_{2\bz}))$, $\bz_{2}=\bA\bx_{2}$, and $\bx_{2} \sim \cN_{\bbC}(\bx_{2};\br_{2\bx},\mathrm{Diag}(\bv_{2\bx}))$. From lines 7 and 17 in Algorithm 1, we know that only diagonal elements of $\bQ_{2\bx}$ should be transmitted to module B. We obtain the SVD of matrix A as follows,
\begin{equation}\label{SVDA}
\bA=\bU\mathbf{\Sigma}\bV,
\end{equation}
where $S_{i}$ is the $i$th nonzero element of $\mathbf{\Sigma}$, $\lambda_{i}$ is the $i$th eigenvalue of $\bA^{H}\bA$ which $\lambda_{i}=S_{i}^{2}$. Therefore, $\bQ_{2\bx}$ is rewritten as
\begin{equation}\label{Q2xSVD}
  \bQ_{2\bx} = \bV(\mathrm{Diag}(\mathbf{1}\oslash\bv_{2\bx})+\mathbf{\Sigma}^{H}\mathrm{Diag}(\mathbf{1}\oslash\bv_{2\bz})\mathbf{\Sigma})^{-1}\bV^{H}.
\end{equation}
The diagonal elements of $\bQ_{2\bx}$ are obtained by
\begin{equation}\label{qx}
 q_{x}=\sfE\Bigg\{\frac{1}{\lambda_{i}\gamma_{2z}+\gamma_{2x}}\Bigg\}=\frac{1}{N}\sum_{i=1}^{N}\left({\frac{1}{v_{2x}}+\frac{s_{i}^{2}}{v_{2z}}}\right)^{-1}.
\end{equation}

%
In module B, we perform MMSE estimate of $\bx$ under an AWGN observation
\begin{equation}\label{AWGN_estimate1}
\br_{1\bx}=\bx+\bw_{1\bx},
\end{equation}
and average MSE of the estimate is $\mathsf{MSE}_{\rmx}$, which is given by
\begin{equation}\label{MSEX1}
   \mathsf{MSE}_{\rmx}=\sfE\{|x-\hat{x}|^{2}\}.
\end{equation}

After obtaining the extrinsic information $\bv_{2\bx}$ and $\bv_{2\bz}$ from module B, module C performs the LMMSE estimate under
the restriction, $\bz_{2}=\bA\bx_{2}$ which is the same as (\ref{LMMSE_estimate1}). The diagonal elements of $\bQ_{2\bz}$ can be derived using a similar method as (\ref{Q2xSVD}), which is given by

\begin{equation}\label{qxqz}
   q_{z}=\sfE\Bigg\{\frac{\lambda_{i}}{\lambda_{i}\gamma_{2z}+\gamma_{2x}}\Bigg\}=\frac{1}{M}\sum_{i=1}^{N}s_{i}^{2}\left({\frac{1}{v_{2x}}+\frac{s_{i}^{2}}{v_{2z}}}\right)^{-1}.
\end{equation}

By combining Eqs. (\ref{v1zpost2})-(\ref{VAR_mixed}), (\ref{qx}), (\ref{MSEX1})-(\ref{qxqz}), and equations of computing extrinsic information, we can derive whole SE equations for Proposition \ref{proposition1}.

\section*{Appendix B: Derivation of the Saddle-point of $\mathcal{F}$}\label{AP2}

In this appendix, we adopt the replica method in the field of statistical physics to calculate $\mathcal{F}$ in the large-system limit and derive its saddle points. From \cite{Nishimori2001Statistical}, free energy $\mathcal{F}$ is given by
\begin{equation}\label{eq:LimF}
        \mathcal{F} = -\frac{1}{M} \lim_{\tau\to 0}\frac{\partial}{\partial \tau}\log \sfE\left[ \sfP^{\tau}(\tilde{\by})\right].
    \end{equation}

    We define the likelihood distribution of the received signals under (\ref{system_model}) conditional on the unknown parameters as follows:
    \begin{equation} \label{likelihood_delta}
        \sfP(\tilde{\by}|\bx) \triangleq \prod_{j=1}^{N} \int \rmd  z_{j} \sfP_{\mathrm{out}}(\tilde{y}_{j}\mid z_j)
        \delta{\left( z_{j} - \ba_{j}^H \bs \right)},
    \end{equation}
\begin{figure*}
    \begin{subequations} \label{defsf_E1}
        \begin{align}
            \cG^{(\tau)}(\bQ_{x},\bQ_{w}) &=  \frac{1}{M} \log {\sf E}_{\bA} {\left[ \prod_{a=1}^{\tau} e^{-\sfj \bw^{(a) H} \bA \bx^{(a)} - \sfj (\bA \bx^{(a)})^{H}\bw^{(a)}  } \right]}, \label{defG} \displaybreak[0]\\
            \mu^{(\tau)}(\bQ_{x}) &= {\sf E}_{\bX}{\left[\int \prod_{1\leq a \leq b }^{\tau}\delta{\left( {\left(\bx^{(a)}\right)}^H\bx^{(b)} -M[\bQ_{x}]_{a,b}\right)} \rmd  [\bQ_{x}]_{a,b} \right]}, \label{defmux} \displaybreak[0]\\
            \mu^{(\tau)}(\bQ_{w}) &= \int{ \rmd\tilde{\by}} \int{ \rmd\bZ } \int{ \rmd\bW }
            {\left( \int \prod_{1\leq a \leq b }^{\tau}\delta{\left( {\left(\bw^{(a)}\right)}^H\bw^{(b)} -N[\bQ_{w}]_{a,b}\right)} \rmd  [\bQ_{w}]_{a,b} \right)} \nonumber \displaybreak[0]\\
            & \hspace{1cm} \times \prod_{a=1}^{\tau} {\sfP_\mathrm{out}{\left( \tilde{\by}\mid \bz^{(a)}\right)}} e^{-\sfj \bw^{(a)H} \bz^{(a)} -\sfj \bz^{(a)H} \bw^{(a)} }. \label{defmuw}
        \end{align}
    \end{subequations}
    \noindent\rule[0.25\baselineskip]{\textwidth}{0.1pt}
    \end{figure*}
 where $\delta(\cdot)$ denotes Dirac's delta. We know that $\sfP(\tilde{\by}) = \int_{\bx}\sfP(\tilde{\by}|\bx)\sfP(\bx)\rmd\bx$ is marginal likelihood. Using the Fourier representation of the $\delta$  via auxiliary variables $\bw=[ w_m ] \in \bbC^{N}$ to (\ref{likelihood_delta}),
    and through the replica method, we compute the replicate partition function $\sfE\left[ \sfP^{\tau}(\by)\right]$ given by
  \begin{align}\label{replica_partition_function}
        \sfE_{\tilde{\by}}\left[\sfP^{\tau}(\tilde{\by})\right]
      & = \int{\rmd\tilde{\by}} ~{\sf E}_{\bA,\bX}\Bigg[ \int{ \rmd\bZ } \int{ \rmd\bW } \times    \nonumber \\
      & \left( \prod_{a=1}^{\tau} \sfP_\mathrm{out}{\left( \tilde{\by}\Big| \bz^{(a)}\right)} e^{-\sfj \bw^{(a)H} \bz^{(a)} -\sfj \bz^{(a)H} \bw^{(a)} } \right) \times \nonumber \\
      & \left( \prod_{a=1}^{\tau} e^{\sfj \bw^{(a) H} \bA \bx^{(a)} + \sfj (\bA \bx^{(a)})^{H} \bw^{(a)} } \right) \Bigg],
  \end{align}

    where $\bz^{(a)}$ and $\bx^{(a)}$ are the $a$-th replica of $\bz$ and $\bx$, respectively; and $\bZ \triangleq \{
    \bz^{(a)}, \forall a \}$, $\bW \triangleq \{ \bw^{(a)}, \forall a \}$, $\bX \triangleq \{ \bx^{(a)}, \forall a \}$. Here, $\{\bx^{(a)}\}$ are random vectors taken
    from the distribution $\sfP(\bx) $ for $a=1, \dots, \tau$. In addition, $\int\rmd \tilde{\by}$ denotes the
    integral w.r.t.~a discrete measure because the quantized output $\tilde{\by}$ is a finite set.

    To evaluate the expectation w.r.t. $\bA$ and $\bX$ in (\ref{replica_partition_function}), we introduce two $\tau \times \tau$ matrices $\bQ_{x}$ and $\bQ_{w}$ whose elements are
    defined by $[\bQ_{x}]_{a,b}\triangleq\frac{1}{M}\left(\bx^{(a)}\right)^H\bx^{(b)}$ and $[\bQ_{w}]_{a,b}\triangleq\frac{1}{N}\left(\bw^{(a)}\right)^H\bw^{(b)}$. The definitions
    of $\bQ_{x}$ and $\bQ_{w}$ are equivalent to
    \begin{align}
        1 &= \int \prod_{1\leq a \leq b }^{\tau}\delta{\left( {\left(\bx^{(a)}\right)}^H\bx^{(b)} -M[\bQ_{x}]_{a,b}\right)} \rmd  [\bQ_{x}]_{a,b} \\
        1 &= \int \prod_{1\leq a \leq b }^{\tau}\delta{\left( {\left(\bw^{(a)}\right)}^H\bw^{(b)} -N[\bQ_{w}]_{a,b}\right)} \rmd  [\bQ_{w}]_{a,b},
    \end{align}
    where $\delta(\cdot)$ denotes Dirac's delta. Inserting the above expressions into (\ref{replica_partition_function}) yields
    \begin{equation} \label{E1}
        \sfE_{\tilde{\by}}\left[ \sfP^{\tau}(\tilde{\by})\right]
        = \int e^{N \cG^{(\tau)}(\bQ_{x},\bQ_{w})} \rmd \mu^{(\tau)}(\bQ_{x}) \rmd \mu^{(\tau)}(\bQ_{w}),
    \end{equation}

    where $\cG^{(\tau)}(\bQ_{x},\bQ_{w})$, $\mu^{(\tau)}(\bQ_{x})$, and $\mu^{(\tau)}(\bQ_{w})$ are given by (\ref{defsf_E1}) at the top of this page. 
    We notice that by introducing the $\delta$-functions, the expectations over $\bX$ can be separated to an expectation over all possible
    covariance $\bQ_{x}$ and all possible $\bX$ configurations w.r.t. a prescribed set of $\bQ_{x}$. Therefore, we can separate the expectations over $\bA$ and $\bX$ respectively in (\ref{defG}) and (\ref{defmux}).
    A similar concept applies to separating the expectations over $\bA$ and $\bW$.
    Using similar strategy in \cite{wang2017Gturbo}, we can calculate each term of (\ref{defsf_E1}). 

    First, we evaluate $\cG^{\tau}(\bQ_{x},\bQ_{w})$ by noticing
    \begin{align} \label{eq:E_uAx1}
    &{\sf E}_{\bA} {\left[ \prod_{a=1}^{\tau} e^{-\sfj \bw^{(a) H} \bA \bx^{(a)} - \sfj (\bA \bx^{(a)})^{H}\bw^{(a)}  } \right]} =  \\
    & {\sf E}_{\bA} {\left[ e^{-\sfj \sum_{a=1}^{n} \tilde{\bw}^{(a)H} \mathbf{\Lambda} \tilde{\bx}^{(a)} + \tilde{\bx}^{(a)H} \mathbf{\Lambda} \tilde{\bw}^{(a)} } \right]},
    \end{align}
    where $\tilde{\bw}^{(a)}=\bU^{H}\bw^{(a)}$ and $\tilde{\bx}^{(a)} =  \bV^{H}\bx^{(a)}$. Because $\bU$ and $\bV$ are unitary matrices, the covariances of $(\tilde{\bx}^{(a)},\tilde{\bx}^{(b)})$ and $(\tilde{\bw}^{(a)},\tilde{\bw}^{(b)})$ are given by the following:
    \begin{align}
    &\frac{1}{M}\left(\tilde{\bx}^{(a)}\right)^H\tilde{\bx}^{(b)} = \frac{1}{M}\left(\bx^{(a)}\right)^H\bx^{(b)} = [\bQ_{x}]_{a,b}, \label{defQx}\\
    &\frac{1}{N}\left(\tilde{\bw}^{(a)}\right)^H\tilde{\bw}^{(b)} = \frac{1}{N}\left(\bw^{(a)}\right)^H\bw^{(b)} = [\bQ_{w}]_{a,b}. \label{defQw}
    \end{align}

    Notice that the dependence on the replica indices does not affect the physics of the system because replicas have been introduced artificially. Assuming {\it replica symmetry} (RS), i.e.,
    \begin{equation}
    \left\{\begin{aligned}
    \bQ_{x}&=c_{x} \bI_{\tau}+q_{x}{\bf 11}^H,\\
    \bQ_{w}&=c_{w} \bI_{\tau}+q_{w}{\bf 11}^H,
    \end{aligned}\right.
    \end{equation}
    therefore seems natural.
    With the RS, we can obtain the following:
    \begin{equation} \label{E_uAx2}
    \cG^{(\tau)}(\bQ_{x},\bQ_{w})
    =  (\tau-1)G(c_{x},c_{w}) + G(c_{x}+\tau q_{x},c_{w}+\tau q_{w}),
    \end{equation}
    where
    \begin{align} \label{E_uAx3}
    \small
      & G(x,u)= \mathsf{Extr}_{\chi_{x},\chi_{w}} \Big\{ \chi_{x} x + \alpha \chi_{w}u- (1-\alpha)\log(\chi_{x})-    \\ \nonumber
      & \alpha\sfE_{\lambda}\log(\chi_{x}\chi_{w} + \lambda ) \Big\} -\log x -\alpha\log u - \alpha-1,
    \end{align}
  and $\mathsf{Extr}_{x}\{ f(x) \}$ denotes the extreme value of $f(x)$ w.r.t.~$x$.

 \begin{figure*}
    \begin{align}\label{LT_Rx2}
    \small
        \frac{1}{N}\log{\sf E}_{\tilde{\bX}}{\left[e^{\mathsf{tr}\left(\tilde{\bQ}_{x} \tilde{\bX}^H\tilde{\bX}\right)}\right]}=\int \rmd\bu_{x}
        \left( {\sf E}_{\tilde{\bx}}{\left[e^{ -\| \bu_{x} - \sqrt{\tilde{q}_{x}}\tilde{\bx}\|^2 + (\tilde{q}_{x}-\tilde{c}_{x})\tilde{\bx}^{H}\tilde{\bx} }\right]}\right)
        \left( {\sf E}_{\tilde{\bx}}{\left[e^{\left(\sqrt{\tilde{q}_{x}}\tilde{\bx}\right)^H \bu_{x}+\bu_{x}^H\sqrt{\tilde{q}_{x}}\tilde{\bx}-\tilde{c}_{x}\tilde{\bx}^H \tilde{\bx}}\right]} \right)^{\tau-1}.
    \end{align}

    \begin{align}\label{de}
    \cR_{w1}(\tilde\bQ_{w}) \triangleq \int{ \rmd\tilde{\by} } \int{ \rmd\bZ } \int{ \rmd\bW } \Bigg(
    \prod_{a=1}^{\tau} {\mathsf{P}_\mathrm{out}{\left( \tilde{\by} \mid \bz^{(a)}\right)}}
    \times e^{-\sfj \bw^{(a)H} \bz^{(a)} -\sfj \bz^{(a)H} \bw^{(a)} } \Bigg)
    e^{\mathsf{tr}\left(\tilde\bQ_{w} \bW^H\bW\right)}.
    \end{align}
 \begin{align}\label{Rw2}
     \cR_{w1}(\tilde\bQ_{w})&= \int \rmD\bu_{w} \Bigg( \prod_{a=1}^{\tau} \int \rmd\bz^{(a)} \int \rmd\bw^{(a)} {\mathsf{P}_\mathrm{out}{\left( \tilde{\by} \mid \bz^{(a)}\right)}}\times e^{-\sfj \bw^{(a)H} \bz^{(a)} -\sfj \bz^{(a)H} \bw^{(a)} } \Bigg) \nonumber \\
    & ~~ \times  e^{ (\sum_{a} \sfj \sqrt{\tilde{q}_{w}}\bw^{(a)})^{H} \bu_{w}
        + \bu_{w}^{H} (\sum_{a} \sfj \sqrt{\tilde{q}_{w}}\bw^{(a)})  - \sum_{a}\tilde{c}_{w}\bw^{(a)H} \bw^{(a)} } \nonumber \\
    &= \int \rmD\bu_{w} \left( \int \rmD\bv_{w} \, {\mathsf{P}_\mathrm{out}{\left( \tilde{\by} \Big| \sqrt{\tilde{c}_{w}} \bv_{w} + \sqrt{\tilde{q}_{w}}\bu_{w} \right)}}  \right)^{\tau}.
    \end{align}
\begin{align} \label{FinaRw}
        \left. \partial\cR_{w}^{(\tau)}(\bQ_{w}) /\partial\tau \right|_{\tau=0} &= \alpha \max_{\tilde{c}_{w},\tilde{q}_{w}}\Bigg\{
        \sum_{\tilde{y}} \, \rmD u_{w} {\left( \int \rmD v_{w} \mathsf{P}_\mathrm{out}{\left( \tilde{y} \Big| \sqrt{\tilde{c}_{w}} v_{w} + \sqrt{ \tilde{q}_{w}} u_{w}  \right)} \right)}\nonumber \\
        &\times \log   {\left( \int \rmD v_{w} \mathrm{P}_{\sf out}{\left( q \Big| \sqrt{\tilde{c}_{w}} v_{w} + \sqrt{ \tilde{q}_{w}} u_{w}  \right)} \right)} - \tilde{c}_{w} (q_{w}+c_{w}) + \tilde{q}_{w} c_{w}
        \Bigg\}.
\end{align}
    \noindent\rule[0.25\baselineskip]{\textwidth}{0.1pt}
\end{figure*}

    Next, we consider $\mu^{(\tau)}(\bQ_{x})$ in (\ref{defmux}). $\mu^{(\tau)}(\bQ_{x}) = e^{N \cR_{x}^{(\tau)}(\bQ_{x})+\cO(1)}$, where
    $\cR_{x}^{(\tau)}(\bQ_{x})$ is the rate measure of $\mu^{(\tau)}(\bQ_{x})$ and is given by \cite{Wen2007IT}
    \begin{equation} \label{Rx1}
    \small
        \cR_{x}^{(\tau)}(\bQ_{x}) = \max_{\tilde{\bQ}_{x}}\left\{
        \frac{1}{N}\log{\sf E}_{\tilde{\bX}}{\left\{e^{\mathsf{tr}\left(\tilde{\bQ}_{x} \tilde{\bX}^H\tilde{\bX}\right)}\right\}}
        - \mathsf{tr}{\left(\tilde{\bQ}_{x} \bQ_{x}\right)} \right\}
    \end{equation}
    with $\tilde{\bQ}_{x}\in {\mathbb C}^{\tau\times\tau}$ being a symmetric matrix.
    Furthermore, we assume the RS, i.e., $\tilde{\bQ}_{x} =  \tilde{q}_{x} {\bf 11}^H - \tilde{c}_{x} \bI_{\tau}$.
    By assuming the RS, using the Hubbard-Stratonovich transformation and introducing the auxiliary vector
    $\bu_{x} \in \mathbb{C}^{N}$, the first term of (\ref{Rx1}) can be written as follows:

    With the RS assumption, the last term of (\ref{Rx1}) can now be expressed as (\ref{LT_Rx2}) at the top of this page.
    \begin{equation} \label{RSQQ}
    \mathsf{tr}{\left(\tilde{\bQ}_{x}\bQ_{x}\right)}
    = (-\tilde{c}_{x}+\tau\tilde{q}_{x})(c_{x}+\tau q_{x}) - (\tau-1) \tilde{c}_{x}c_{x}.
    \end{equation}
    Substituting (\ref{LT_Rx2}) and (\ref{RSQQ}) into (\ref{Rx1}) and taking the derivative w.r.t. $\tau$ at $\tau = 0$, we obtain the following:
    \begin{align} \label{FinaRx}
       & \left. \partial\cR_{x}^{(\tau)}(\bQ_{x})/\partial\tau \right|_{\tau=0} = \\ \nonumber
        &\max_{\tilde{c}_{x},\tilde{q}_{x}}\Bigg\{
        \int \rmD u_{x} {\sf E}_{\tilde{x}}{\left[ e^{- |u_{x}-\sqrt{\tilde{q}_{x}}\tilde{x}|^2 + (\tilde{q}_{x}-\tilde{c}_{x}) |\tilde{x}|^2 }   \right]}\\ \nonumber
       & \times \log{\sf E}_{x}{\left[ e^{-\tilde{c}_{x}|\tilde{x}|^2 + {\rm Re}\left[\sqrt{\tilde{q}_{x}} u_{x}^{*} \tilde{s} \right] }   \right]}
        - \tilde{c}_{x} (c_{x}+q_{x}) + \tilde{q}_{x} c_{x} \Bigg\}.
    \end{align}

    Similarly, we calculate $\mu^{(\tau)}(\bQ_{w})$ in (\ref{defmuw}) and assume the RS $\tilde{\bQ}_{w} =  -\tilde{q}_{w} {\bf 11}^H - \tilde{c}_{w} \bI_{\tau}$.
    $\mu^{(\tau)}(\bQ_{w}) = e^{M \cR_{w}^{(\tau)}(\bQ_{w})+\cO(1)}$, where $\cR_{w}^{(\tau)}(\bQ_{w})$ is the rate measure of $\mu^{(\tau)}(\bQ_{w})$ and is
    given by the following:
    \begin{equation} \label{Rw1}
    \cR_{w}^{(\tau)}(\bQ_{w})= \max_{\tilde{\bQ}_{w}}\left\{
    \frac{1}{N}\log\cR_{w1}(\tilde\bQ_{w})
    - \mathsf{tr}\left(\tilde{\bQ}_{w} \bQ_{w}\right) \right\},
    \end{equation}
    where we define (\ref{de}).

    By using the Hubbard-Stratonovich transformation and introducing the auxiliary
    vector $\bu_{w} \in \mathbb{C}^{N}$, we obtain (\ref{Rw2}).
   where the last equality follows the facts that $\bv_{w} \triangleq \frac{1}{\sqrt{\tilde{c}_{w}}} \left(\sqrt{\tilde{q}_{w}}\bu_{w} - \bz \right)$ and $\rmD\bv_{w} = \frac{1}{\pi^{N}} e^{-\bv_{w}^H\bv_{w}} $.
    With the RS assumption, the last term of (\ref{Rw1}) can now be expressed as follows:
    \begin{equation} \label{RSQQw}
    \mathsf{tr}{\left(\tilde\bQ_{w}\bQ_{w}\right)}
    = (-\tilde{c}_{w}+\tau\tilde{q}_{w})(c_{w}+\tau q_{w}) - (\tau-1) \tilde{c}_{w}c_{w}.
    \end{equation}
    Substituting (\ref{Rw2}) and (\ref{RSQQw}) into (\ref{Rw1}) and taking the derivative w.r.t. $\tau$ at $\tau = 0$, we obtain the (\ref{FinaRw}) at the top of this page.
%
    The integration over $\bQ$ in (\ref{E1}) can be performed via the saddle point method as $M\rightarrow\infty$,
    which yields the following:
    \begin{align}\label{sadP}
    &\lim_{M\rightarrow\infty}\frac{1}{M} \mathsf{E}_{\tilde{\by}}{\left[ \mathsf{P}^{\tau}(\tilde{\by})\right]}= \nonumber \\ &\max_{\bQ_{x},\bQ_{w}}\Big\{{\cG}^{(\tau)}(\bQ_{x},\bQ_{w})-
    \cR_{x}^{(\tau)}(\bQ_{x}) - \cR_{w}^{(\tau)}(\bQ_{w}) \Big\}
    \end{align}
The extremum over $\{c_{x},c_{w}, \tilde{c}_{x}, \tilde{c}_{w}, q_{x}, q_{w}, \tilde{q}_{x}, \tilde{q}_{w}\}$ can be obtained by equating the corresponding partial derivatives of the RS expression $\Phi$ to zero. With the normalization constraint $\sfE\{\sfP^{\tau}(\tilde{\by})\}=1$, $c_{x}+q_{x}=\sfE\{|X|^{2}\}\triangleq T_{x}$, $c_{w}+q_{w}=0$, $-\tilde{c}_{x}+\tilde{q}_{x}=0$, and $\tilde{c}_{w}+\tilde{q}_{w}=\sfE\{\lambda\}\sfE\{|X|^2\}\triangleq \tilde{T}_{w}$ can be easily obtained. These relationships yield
  \begin{multline}\label{eq:GenFree2}
     \Phi= \mathsf{Extr}_{q_{x},q_{w}}\Bigg\{ G(T_{x}-q_{x},q_{w})  + \alpha q_{w} \tilde{T}_{w}- I\left(x;z\left|\sqrt{\tilde{q}_{x}} \right.\right)\\
    + \tilde{q}_{x} (T_{x}-q_{x}) + \\
    \alpha\sum_{\tilde{y}} \int \rmD v \cP_\mathsf{out}(\tilde{y}| v; \tilde{q}_{w}) \log \cP_\mathsf{out}(\tilde{y}| v;\tilde{q}_{w})
    - \tilde{q}_{w} q_{w} \Bigg\},
  \end{multline}

 \begin{equation}
    \cP_{\mathsf {out}}(\tilde{y}| v; \tilde{q}_{w}) = \int \rmD u \mathrm{P}_{\mathsf{out}}{\left( \tilde{y} \Big| \sqrt{ \tilde{T}_{w} -\tilde{q}_{w}} u + \sqrt{ \tilde{q}_{w}} v  \right)},\\
\end{equation}
 \begin{align}
    I{\left(x;z\left|\sqrt{\tilde{q}_{x}} \right.\right)}
    &= -  \int \mathrm{d}z\sfE_{x}\{e^{-|z-\sqrt{\tilde{q}_{x}}x|^{2}}\}\\
    & \times \log \sfE_{x}\{e^{-|z-\sqrt{\tilde{q}_{x}}x|^{2}}\}- 1,
 \end{align}
  \begin{equation}
    \mathcal{P}_{x}(z\left|x;\sqrt{\tilde{q}_{x}}\right.) = \frac{1}{\pi}e^{-|z-\sqrt{\tilde{q}_{x}}x|^{2}}.
    \end{equation}
    with a defined scalar Gaussian channel
    \begin{equation}\label{eqscalar}
      z=\sqrt{\tilde{q}_{x}}x+w,
    \end{equation}
    where $x \sim \sfP_\sfX(x)$ and $w\sim \cN(0,1)$.
    The saddle-point of (\ref{eq:GenFree2}) can be rewritten as
  \begin{subequations} \label{eq:sdPoint2}
        \begin{align}
        \tilde{q}_{w}&=\tilde{T}_{w}+\chi_w-\frac{1}{q_{w}}, \label{eq:sdPoint2_1}\displaybreak[0]\\
       q_{w}&=\frac{1}{2}\sum_{b=1}^{B}\int \rmD v\frac{\bigg(\Psi_{b}^{'}(\sqrt{\tilde{q}_{w}}v)\bigg)^{2}}{\Psi_{b}(\sqrt{\tilde{q}_{w}}v)}, \label{eq:sdPoint2_2} \displaybreak[0]\\
       \tilde{q}_{x}&=-\chi_x+\frac{1}{T_{x}-q_{x}}, \label{eq:sdPoint2_3} \displaybreak[0]\\
         \chi_{x}&=\frac{1}{\mathsf{MSE}_{\rmx}(\tilde{q}_{x})}-\tilde{q}_{x},\label{eq:sdPoint2_4}
        \end{align}
    \end{subequations}
%
%
%
where
\begin{equation}\label{eq8}
  \Psi_{b}(V_{o})\triangleq \Phi\bigg(\frac{\sqrt{2}r_{b}-V_{o}}{\sqrt{\sigma^{2}+\tilde{T}_{w}-\tilde{q}_{w}}}\bigg)-\Phi\bigg(\frac{\sqrt{2}r_{b-1}-V_{o}}{\sqrt{\sigma^{2}+\tilde{T}_{w}-\tilde{q}_{w}}}\bigg)
\end{equation}
\begin{equation}\label{eq9}
\Psi_{b}^{'}(V_{o})\triangleq \frac{\partial\Psi_{b}(V_{o})}{\partial V_{o}}=\frac{e^{-\frac{(\sqrt{2}r_{b}-V_{o})^{}2}{2(\sigma^{2}+\tilde{T}_{w}-\tilde{q}_{w})}}-e^{-\frac{(\sqrt{2}r_{b}-V_{o})^{}2}{2(\sigma^{2}+\tilde{T}_{w}-\tilde{q}_{w})}}}
{\sqrt{2\pi(\sigma^{2}+\tilde{T}_{w}-\tilde{q}_{w})}}
\end{equation}
In addition, We obtain that the extremum points should satisfy the following equality
\begin{equation}\label{eq10}
  q_{w}=\mathrm{E}\bigg\{\frac{\chi_{x}} {\chi_{x}\chi_{w}+\lambda} \bigg\},
\end{equation}
\begin{equation}\label{eq11}
  T_{x}-q_{x}=(1-\alpha)\frac{1}{\chi_{x}}+\alpha \mathrm{E}\bigg\{ \frac{\chi_{w}}{\chi_{x}\chi_{w}+\lambda}\bigg\},
\end{equation}
where $\alpha=M/N$ denotes measurement ratio. Comparing $(\ref{eq:sdPoint2})$-$(\ref{eq11})$ with Proposition \ref{proposition1}, we can conclude that they share the same SE equations.

\ifCLASSOPTIONcaptionsoff
  \newpage
\fi

\begin{IEEEbiography}[{\includegraphics[width=1in,height=1.25in,clip,keepaspectratio]{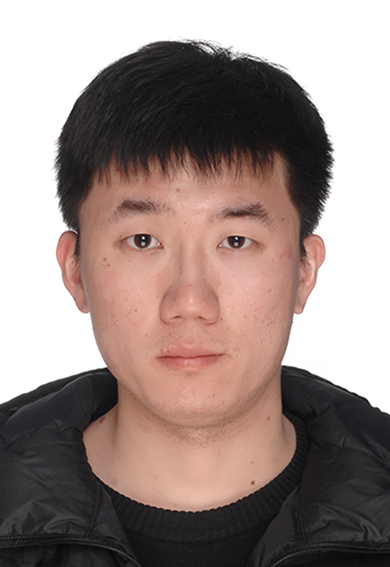}}]
{\textbf{Hengtao He}} (S'15) received the B.S. degree in communications
engineering from Nanjing University of Science
and Technology, Nanjing, China, in 2015. He is
currently working towards the Ph.D. degree in
information and communications
engineering, Southeast University,
China, under the supervision of Prof. Shi Jin.
His areas of interests currently include millimeter wave communications, massive MIMO, and machine learning for wireless communications.

\end{IEEEbiography}

\begin{IEEEbiography}[{\includegraphics[width=1in,height=1.25in,clip,keepaspectratio]{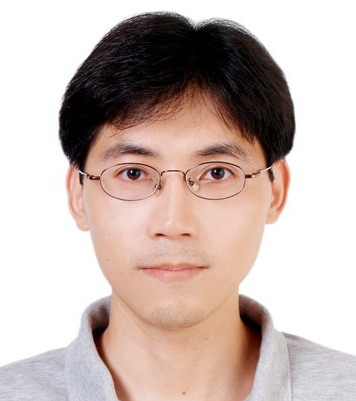}}]
{\textbf{Chao-Kai Wen}} (S'00--M'04) received the Ph.D. degree from the Institute of Communications Engineering, National Tsing Hua University, Taiwan, in 2004. He was with Industrial Technology Research Institute, Hsinchu, Taiwan and MediaTek Inc., Hsinchu, Taiwan, from 2004 to 2009.
Since 2009, he has been with National Sun Yat-sen University, Taiwan, where he is Professor of the Institute of Communications Engineering.
His research interests center around the optimization in wireless multimedia networks.
\end{IEEEbiography}

\begin{IEEEbiography}[{\includegraphics[width=1in,height=1.25in,clip,keepaspectratio]{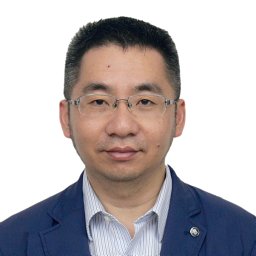}}]
{\textbf{Shi Jin}} (S'06-M'07-SM'17) received the B.S. degree in communications engineering from Guilin University of Electronic Technology, Guilin, China, in 1996, the M.S. degree from Nanjing University of Posts and
Telecommunications, Nanjing, China, in 2003, and the Ph.D. degree in information and communications engineering from the Southeast University, Nanjing, in 2007. From June 2007 to October 2009, he was a Research Fellow with the Adastral Park Research Campus, University College London, London, U.K. He is currently with the faculty of the National Mobile Communications Research Laboratory, Southeast University. His research interests include space time wireless communications, random matrix theory, and information theory. He serves as an Associate Editor for the IEEE Transactions on Wireless Communications, and IEEE Communications Letters, and IET Communications. Dr. Jin and his co-authors have been awarded the 2011 IEEE Communications Society Stephen O. Rice Prize Paper Award in the field of communication theory and a 2010 Young Author Best
Paper Award by the IEEE Signal Processing Society.
\end{IEEEbiography}

\end{document}